\documentclass[aps,prb%,draft,twocolumn
,groupedaddress,showpacs,nofootinbib,10pt]{revtex4-1}
\DeclareMathAlphabet{\mathpzc}{OT1}{pzc}{m}{it}
\usepackage{xcolor}
\usepackage{braket}
\usepackage{slashed}
\usepackage{mathtools}
\usepackage{nicefrac}
\usepackage{amssymb}
\usepackage{mathrsfs}
\usepackage{IEEEtrantools}
\definecolor{BrickRed}{cmyk}{0, .89, .94, .28} 

\newcommand{\be}{\begin{equation}}
\newcommand{\ee}{\end{equation}}
\newcommand{\bea}{\begin{IEEEeqnarray}{rCl}}
\newcommand{\eea}{\end{IEEEeqnarray}}

\newcommand{\ba}{\begin{array}}
\newcommand{\ea}{\end{array}}

\definecolor{lightGray}{RGB}{220,220,220}

\definecolor{MyDarkBlue}{rgb}{0,0.08,0.45}

\definecolor{MyDarkGray}{RGB}{140,140,140}

\newcommand{\bes}{\begin{equation}\begin{split}}
\newcommand{\ees}{\end{split}\end{equation}}

\definecolor{lightGray}{RGB}{220,220,220}
\definecolor{MyDarkGray}{RGB}{140,140,140}

\definecolor{MyDarkBlue}{rgb}{0,0.08,0.45}

\newlength{\eqboxstorage}

\newcommand{\ed}{\end{document}}

\newcommand{\bl}{\begin{list}{$\circ$}{}}
\newcommand{\el}{\end{list}}

\definecolor{cadmiumgreen}{rgb}{0.0, 0.42, 0.24}

\usepackage{graphicx}% Include figure files
\usepackage{dcolumn}% Align table columns on decimal point
\usepackage{bm}% bold math
\usepackage{xcolor}
\usepackage{hyperref}
\newcommand{\pade}{Pad\'e~}
\newcommand{\Oc}{\mathcal{O}}
\newcommand{\Un}{\mathcal{U}}

\usepackage{graphicx}
\usepackage{tikz}
\usetikzlibrary{decorations.markings}
\newcommand{\Int}{\int\!\!\!\!\!\!\!
\begin{tikzpicture}[baseline=-0.65ex,scale=0.09,decoration={markings,
mark=at position 0.5cm with {\arrow[line width=1pt]{>}},
mark=at position 2cm with {\arrow[line width=1pt]{>}},
mark=at position 7.85cm with {\arrow[line width=1pt]{>}},
mark=at position 9cm with {\arrow[line width=1pt]{>}}
}
]
\path[draw,line width=0.6pt,postaction=decorate] (-1,0) -- (2,0) arc (0:180:2) -- (-1,0);
\end{tikzpicture}\;}

\begin{document}

\title{Diagonal Pad\'e Approximant of the one-body Green's function,
a study on Hubbard rings}

\author{Walter Tarantino}
\email{walter.tarantino@mat.ethz.ch}
\affiliation{Materials Theory, ETH Z\"urich, Wolfgang-Pauli-Strasse 27, CH-8093 Z\"urich, Switzerland}
\affiliation{Laboratoire des Solides Irradi\'es, Ecole Polytechnique, CNRS, CEA, Universit\'e Paris-Saclay, F-91128 Palaiseau, France}
\author{Stefano Di Sabatino}
%\author{Lucia Reining}
\affiliation{Laboratoire des Solides Irradi\'es, Ecole Polytechnique, CNRS, CEA, Universit\'e Paris-Saclay, F-91128 Palaiseau, France}

%\date{\today}

%%%%%%%%%%%%%%%%%%%%%%%%%%%%%%%%%%%%%%%%%%%%%%%%%%%%%%%%%%%%%%%%%%%%%%%%%%%%%%%%%%%%%%%
\begin{abstract}
\pade approximants to the many-body Green's function
can be build by rearranging terms of its perturbative expansion.
The hypothesis that the best use of a finite number of terms of such an expansion
is given by the subclass of \emph{diagonal} \pade approximants
is here tested, and largely confirmed, on a solvable model system, 
namely the Hubbard ring for a variety of site numbers, 
fillings and interaction strengths.
\end{abstract}
%%%%%%%%%%%%%%%%%%%%%%%%%%%%%%%%%%%%%%%%%%%%%%%%%%%%%%%%%%%%%%%%%%%%%%%%%%%%%%%%%%%%%%%
%\keywords{}
%%%%%%%%%%%%%%%%%%%%%%%%%%%%%%%%%%%%%%%%%%%%%%%%%%%%%%%%%%%%%%%%%%%%%%%%%%%%%%%%%%%%%%%

\maketitle
\section{Introduction}

In the context of quantum field theory, perturbation
theory allows to derive a formal series for the interacting
one-body Green's function in terms of the corresponding
non-interacting one. In an oversimplified notation this
reads
\be\label{pert}
G = G_0 + G_0vG_0G_0 + G_0vG_0G_0vG_0G_0 + ...
\ee
where $v$ denotes the interaction between the particles.
Convergence properties of this series depend on the specific
Hamiltonian one starts from; for most interesting
cases the series is suspected to be, however, either divergent
or merely slowly convergent.
A way to overcome this limit is to sum up an infinite subset of terms,
which is commonly achieved by recasting the calculation of $G$
in the form of a Dyson equation: 
\be\label{dyson}
G = G_0+G_0\Sigma G,
\ee
where $\Sigma$, the so called self-energy of the system,
has itself a (simpler) perturbative expansion in terms of $G_0$ and $v$,
which leads to the following expansion of the Green's function:
\be\label{dysonpert}
G=(G_0^{-1}-vG_0 - vG_0vG_0G_0-...)^{-1}.
\ee
A finite number of terms of the expansion of $\Sigma$ results
in an infinity of terms for the corresponding approximate $G$ and 
it usually greatly improves over the approximation provided by \eqref{pert}. 
Such an expansion is at the heart of state-of-the-art methods used nowadays
for the estimate of $G$ for real systems \cite{reining}.

One way to explain this success is to interpret the approach in terms of 
\pade approximants \cite{bakergraves}.
Given a certain function $f(z)$, if we only know the first $n+1$ coefficients of its Taylor expansion
$f(z)=c_0+c_1 z+c_2 z^2+...+c_n z^n+...$,
we can approximate the function with a power series that gives rise to the same expansion,
which is trivially the $n^{th}$ order truncation of the Taylor series itself,
$f(z)\approx c_0+c_1 z+c_2 z^2+...+c_n z^n$.
This is what standard perturbation theory does.
Alternatively, we can use the rational function
\be\label{padeLM}
\frac{a_0+a_1 z+a_2 z^2+...+a_p z^p}{1+b_1 z+b_2 z^2+...+b_q z^q}
\ee
whose coefficients are determined by the same condition, namely that \eqref{padeLM} 
gives rise to the same Taylor expansion of $f(z)$ up to the $n^{th}$ order.
For given $n$ this criterion identifies not just one,
but a set of rational functions obtained for different values of $p$ and $q$
with the constrain $p+q=n$; these functions are called
`\pade approximants' of order $p$ and $q$ and denoted by $[p/q]_f(z)$, or, in short, $[p/q]$.
The special case $p=q$ is called \emph{diagonal} \pade approximant.
The approximant $[n/0]$ is the truncated Taylor series itself, 
hence the series of \pade approximants can be considered as a way 
to generalise the Taylor series.
For many functions, \pade approximants are found to represent 
better approximations than truncations of the Taylor series \cite{bender}.
Perhaps not surprisingly, this is particularly true when the series diverges:
a divergence in the Taylor expansion of a function reflects the presence of singularities 
over the complex plane $z\in \mathbf{Z}$ which a power series representation is incapable to describe;
a rational function, on the other hand, has singularities of its own which
can be `tuned' to fit those of the function to approximate.
If the Green's function was simply a function of $v$,
the series generated by perturbation theory \eqref{pert} would amount to its Taylor expansion,
or, in terms of \pade approximants, $[n/0]$,
while solving the Dyson equation with the $n^{th}$ truncation of the perturbative expansion of $\Sigma$
would represent a \pade approximant $[0/n]$
(the fact that the Green's function in facts depends on spin/space/time variables
makes the definition of \pade approximants less obvious than its Taylor expansion, 
as we shall later see).
From this perspective, it is then natural to expect that, if the perturbative series for the actual Green's function seems to diverge,
solving the Dyson equation with a perturbative self-energy may then give rise to a sequence of approximations
with better convergence properties, as it turns out to be in many cases of interest.

In fact, one can go one step farther.
As mentioned, for a given order of truncation of the Taylor expansion 
there is a variety of \pade approximants,
according to the order of the numerator and denominator.
For instance, given the $4^{th}$ order one can build
$[4/0]$ (the truncated Taylor expansion itself), 
$[0/4]$ (equivalent to take $\Sigma$ up to fourth order) 
but also $[1/3]$, $[2/2]$, and $[3/1]$. 
A priori, there is no way to know which one would provide the best approximation.
A large amount of evidence from numerical experiments and various applications in physics
had led, however, to conjecture that, whenever no information on the function to approximate other
than its truncated Taylor expansion is given, the best approximation is most likely obtained 
by the \emph{diagonal approximant} $[n/n]$ \cite{bakergraves}.
This rises the question: would the equivalent of a diagonal \pade approximant work better than 
\eqref{dysonpert} in approximating the Green's function?

Obviously, such a question is too vague to admit a neat answer. 
`What are the criteria to say that an approximation is ``better'' than another?',
`what fields and interactions are we considering and what is the value of the coupling constants?'
are just examples of questions one should first clarify.
Moreover, even if the question were formulated with sufficient precision,
practical obstacles, such as the increasing complexity of terms of perturbation theory
and the scarcity of exactly known Green's functions to use as test bench,
would set severe limits to our capability to give a definite answer.
Nonetheless, it does make sense to consider the possibility that diagonal \pade approximants 
may be a valuable tool for developing approximations
of the Green's function of real systems,
for which a very limited number of terms of the expansion \eqref{pert} are computationally accessible
and one wishes to make the best use of them.
In this paper we report a preliminary study on such a possibility.

We here focus on the simplest, nontrivial diagonal approximant, namely $[1/1]$,
and perform a systematic study on a set of models, zero-temperature Hubbard rings,
that are numerically solvable, in the attempt to understand whether the diagonal \pade
may offer a better approximation than \eqref{pert} or \eqref{dysonpert}.
More specifically, in Section \ref{sec:theory}, we set the theoretical framework:
we first derive a general expression of such an approximant for the Green's function
in terms of the expansion of the self-energy; we then simplify that expression in the case
of a two-body interaction for zero temperature
Green's function in the Lehmann representation projected on a basis;
finally we explain how these formulas are modified to avoid double-counting when the starting
point of the expansion $G_0$ is not the completely free Green's function but comes from a
mean field approximation, like Hartree or density functional theory (DFT).
We therefore arrive to an expression for $[1/1]$ for the Green's function 
that can be readily used for model as well as for real systems.
In Section \ref{sec:comparison}, we perform a systematic study of this approximation 
on zero-temperature Hubbard rings. We compare the relative error 
of the approximate spectral functions arising from the three approaches: 
straightforward perturbation theory of eqn. \eqref{pert} (G-PT, henceforth), 
perturbation theory applied to the Dyson equation \eqref{dysonpert} ($\Sigma-$PT)
and the set of diagonal \pade approximants (dP-PT), 
for a variety of number of sites ($L=2,4,6,8,10$),
fillings ($N=2,...,2L-2$) and strengths of interaction ($U/t=0.1,1,10$).
This allows us to attempt to extrapolate
the convergence properties of the three series.
We then focus on the direct comparison between $\Sigma$-PT and dP-PT 
and, finally, we discuss the advantages
of building the series on top of a mean-field approximation.
%In the study of the electronic structure of extended systems,
%perturbation theory only serves as basic framework for more sophisticated methods,
%such as GW and Dynamical Mean Field Theory, that nowadays represent 
%state-of-the-art tools for numerical calculations of the Green's function.
%It is then natural to wonder if dP-PT could represent a competitive alternative to those methods.
%In view of this, in Section \ref{sec:gw} we compare the performance of $P_{[1/1]}[G_0]$ 
%with the GW approximation on the same model.
Conclusions are drawn at the end.

%%%%%%%%%%%%%%%%%%%%%%%%%%%%%%%%%%%%%%%%%%%%%%%%%%%%%%%%%%%%%%%%%%%%%%%%%%%%%%%%%%%%%%%%%%
%%%%%%%%%%%%%%%%%%%%%%%%%%%%%%%%%%%%%%%%%%%%%%%%%%%%%%%%%%%%%%%%%%%%%%%%%%%%%%%%%%%%%%%%%%
%%%%%%%%%%%%%%%%%%%%%%%%%%%%%%%%%%%%%%%%%%%%%%%%%%%%%%%%%%%%%%%%%%%%%%%%%%%%%%%%%%%%%%%%%%

\section{First order diagonal \pade approximant for Green's functions}\label{sec:theory}

For a scalar function $f(z)$ with Taylor expansion $\sum_{i=0}^\infty c_i z^i$, 
the first (nontrivial) diagonal \pade approximant is the rational function 
(here conveniently arranged in the form of a truncated continued fraction):
\be\label{P11scalar}
[1/1]_f(z)=\frac{c_0}{1-z\frac{c_1c_0^{-1}}{1+z\left(c_1c_0^{-1}-c_2c_1^{-1}\right)}}
\ee
whose Taylor expansion correctly matches that of $f(z)$ for the first $1+1+1=3$ terms: 
$[1/1]_f(z)=c_0+c_1 z+c_2 z^2+...$. In order to generalize this formula to cases
in which the coefficients $c_i$ are not just numbers,
but more complex objects like matrices or functions,
as the one we are interested in, we follow Section 8.2 of \cite{bakergraves}.
As one should expect when dealing with matrices,
the generalization of \eqref{P11scalar} is not unique,
due to the fact that a simple product of two scalar coefficients $c_i c_j$
can correspond to either $C_i C_j$ or $C_j C_i$ if the two matrices $C_i$ and $C_j$ do not commute.
The criterion we shall adopt here is that the resulting series of approximant of the
Green's function  $([0/0],\;[0/1],\;[0/2],...)$ must correspond to the series $\Sigma$-PT
given by \eqref{dysonpert}.
This unambiguously identifies a \pade generalization of standard many-body perturbation theory.
We then proceed as follows.

First, suppose $c_i$ is a $N \times N$ complex matrix; we can define the reciprocal of the series
$\sum_{i=0}^\infty c_i z^i$ as $\sum_{j=0}^\infty d_j z^j$ for which:
\be
d_0=c_0^{-1}\mbox{  and  }d_j=-c_0^{-1}\sum_{i=1}^{j}c_{i}d_{j-i},
\ee
for which we can formally write
\be
\left(\sum_{j=0}^\infty d_j z^j\right)\left(\sum_{i=0}^\infty c_i z^i\right)=\left(\sum_{i=0}^\infty c_i z^i\right)\left(\sum_{j=0}^\infty d_j z^j\right)=\mathbb{I}
\ee
with $\mathbb{I}$ the $N \times N$ identity, which is a well-defined order-by-order equality.
Then, if  we consider the matrix ${\bf G}$ with perturbative expansion in the parameter $\lambda$
\be\label{matrixpert}
{\bf G}={\bf G_0}+\lambda{\bf G_1}+\lambda^2 {\bf G_2}+...
\ee
we can write
\bea
{\bf G}&=&\left(\mathbb{I}+\lambda{\bf G_1 G_0}^{-1}
				+\lambda^2 {\bf G_2G_0}^{-1}+...\right){\bf G_0}\\
       &=&\left(\mathbb{I}-\lambda {\bf G_1 G_0}^{-1}
				+\lambda^2 \left({\bf G_1G_0}^{-1}{\bf G_1 G_0}^{-1}
                	-{\bf G_2G_0}^{-1}\right)+...\right)^{-1}{\bf G_0}\\
        &=&\left(\mathbb{I}-\lambda \left({\bf G_1 G_0}^{-1}
				+\lambda \left({\bf G_1G_0}^{-1}{\bf G_1 G_0}^{-1}
                	-{\bf G_2G_0}^{-1}\right)+...\right)\right)^{-1}{\bf G_0}\\
        &=&\left(\mathbb{I}-\lambda \left({\bf G_0 G_1}^{-1}
				+\lambda \left(\mathbb{I}-{\bf G_0 G_1}^{-1}{\bf G_2 G_1}^{-1}\right)+...\right)^{-1}\right)^{-1}{\bf G_0}\\
\eea
where we inverted series between parenthesis twice;
a truncation of the series in the last line yields
\be\label{p11g}
{\bf G}\approx \left(\mathbb{I}-\lambda \left({\bf G_0 G_1}^{-1}
				+\lambda \left(\mathbb{I}-{\bf G_0 G_1}^{-1}{\bf G_2 G_1}^{-1}\right)\right)^{-1}\right)^{-1}{\bf G_0}%:= P_{[1/1]}[{\bf G_0}]
\ee
which is a generalization of \eqref{P11scalar} for the matrix ${\bf G}$.
If ${\bf G}$ is written in terms of a Dyson equation ${\bf G}={\bf G_0}+{\bf G_0 \Sigma G}$,
with ${\bf \Sigma}=\lambda {\bf \Sigma_1}+\lambda^2 {\bf \Sigma_2}+...$, the above procedure leads to
\be\label{p11sigma}
{\bf \Sigma}\approx \lambda \left( {\bf \Sigma_1}^{-1}-\lambda {\bf \Sigma_1}^{-1}{\bf \Sigma_2 \Sigma_1}^{-1} \right)^{-1}.
\ee
Actual Green's functions depend on two spin-space-time coordinates rather than just
two discrete, finite indices, but the generalization of the above is straightforward:
\be\label{P11GF}
G(1,2)\approx \left(G_0^{-1}(1,2)-\left(\Sigma_1^{-1}(1,2)-\int \!d3 d4\; \Sigma_1^{-1}(1,3)\Sigma_2(3,4)\Sigma_1^{-1}(4,2)\right)^{-1} \right)^{-1}:=P_{[1/1]}[G_0(1,2)]
\ee
or, equivalently,
\be\label{P11SIGMA}
\Sigma(1,2)\approx \left(\Sigma_1^{-1}(1,2)-\int \!d3 d4\; \Sigma_1^{-1}(1,3)\Sigma_2(3,4)\Sigma_1^{-1}(4,2)\right)^{-1}
\ee
where $G$ is the Green's function, $G_0$ its noninteracting counterpart,
$\Sigma$ the self-energy,  
$\Sigma_1$, $\Sigma_2$, being the first terms
of its perturbative expansion of order 1 and 2, and the notation $1,2,...$ 
shorthands their spin/space-time dependence
$(\sigma_1,{\bf r}_1,t_1),(\sigma_2,{\bf r}_2,t_2),...$.
A specific symbol for the approximant $[1/1]$ as functional
of the noninteracting Green's function, $P_{[1/1]}[G_0]$, has been introduced.
More precisely, this is equivalent to $[1/1]_\lambda [G]$ with $\lambda$,
the parameter that in G-PT is formally introduced to derive the equivalent
of \eqref{matrixpert}, set to 1.
This notation emphasizes the fact that we start from a given noninteracting Green's function $G_0$
(which encodes the information of the parameter $\lambda$ via $\lim_{\lambda\to 0} G=G_0 $).
Since in the calculation of electronic structures it is customary to build perturbative
approximations upon mean-field ones, like Hartree, Hartree-Fock or DFT,
%Since the `noninteracting' Green's function does not need to necessarily be a completely free
%Green's function, %which we shall hereby denote by $g$,
this notation allows to clearly specify from which noninteracting Green's function we start from
(for instance $P_{[1/1]}[G_0^{(0)}]$, $P_{[1/1]}[G_{\rm H}]$, 
or $P_{[1/1]}[G_{\rm LDA}]$ for a completely free, a Hartree or 
a DFT Green's function in the local density approximation, respectively).
The choice of $G_0$ also determines the exact forms of $\Sigma_1$ and $\Sigma_2$.
If $G_0$ is the Green's function corresponding to the
Hamiltonian $H_0=T+V_{ext}+V_{m.f.}$, where $T$ is the kinetic term, 
and $V_{ext}$ and $V_{m.f.}$ are the terms
that couple the electronic field to the external and the mean-field potential, 
$v_{ext}(\vec{r})$ and $v_{m.f.}(\vec{r})$ respectively,
$\Sigma_1$ and $\Sigma_2$ are given, diagrammatically, by
\be\label{diagrams}
\Sigma_1=\includegraphics[scale=0.5]{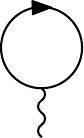}+
\includegraphics[scale=0.5]{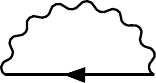}+
\includegraphics[scale=0.8]{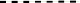}
\;\;\;\mbox{and}\;\;\;
\Sigma_2=
\includegraphics[scale=0.8]{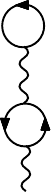}+
\includegraphics[scale=0.8]{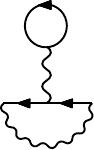}+
\includegraphics[scale=0.8]{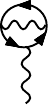}+
\includegraphics[scale=0.8]{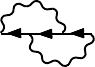}+
\includegraphics[scale=0.8]{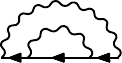}+
\includegraphics[scale=0.8]{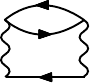}+
\includegraphics[scale=0.6]{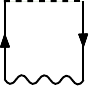}+
\includegraphics[scale=0.6]{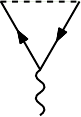}
\ee
where straight lines represent $G_0$, 
wiggly lines the two-body interaction $v(\vec{r}_1,\vec{r}_2)$, and
dashed lines the one-body mean-field potential $v_{m.f.}(\vec{r})$.
Explicit formulas are provided in Appendix \ref{app:diagrams}.

It should be noticed that, in order to have a nontrivial $P_{[1/1]}[G_0]$, 
we must have $\Sigma_1 \neq 0$, otherwise $P_{[1/1]}[G_0]=G_0$.
This is the case, for instance, of
the self-consistent Hartree-Fock Green's function $G_{HF}$, for which
higher order \pade approximants must be considered in order to have a nontrivial
correction to $G_{HF}$.

The Green's function of real systems is often calculated on a truncated orbital basis $\{u_{i\sigma}(\vec{r})\}$ and 
Fourier-transformed to frequency space.
If such a basis diagonalizes the matrix $H_0=T+V_{ext}+V_{m.f.}$,
then $G_0$ can be expressed as a diagonal matrix whose components are 
\be\label{gij}
g_{ij}(\omega)= \delta_{ij}\left\{
\begin{array}{lr}
\frac{1}{\omega-i\eta+(\epsilon_j-\epsilon_0)} & j\in\Oc\\
\frac{1}{\omega+i\eta-(\epsilon_j-\epsilon_0)} & j\in\Un
\end{array}
\right.
\ee
where $\Oc$ denotes the set of occupied orbitals, while $\Un$ the set of unoccupied ones,
$\epsilon_0$ the ground-state energy and $\epsilon_j$
the energy of the $j^{th}$ orbital;
the above diagrammatic expression \eqref{diagrams} then reduces to 
\be\label{sigmasimple1}
\Sigma_{ij}^{(1)}=\sum_{k\in\Oc}(v_{ikkj}-v_{ikjk})-i\delta_{ij}v^{m.f.}_{ii}
\ee
\begin{multline}\label{sigmasimple2}
\Sigma_{ij}^{(2)}(\omega)=\sum_{o,q,r}\Oc_q\left(
    	\frac{\Oc_o \Un_r}{\epsilon_o + \epsilon_r}+
        \frac{\Oc_r \Un_o}{\epsilon_r + \epsilon_o}
    	\right)(v_{iojr}-v_{iorj})(v_{qroq}-v_{qrqo})+\\
    	 +\sum_{nps} v_{isnp}(v_{npjs}-v_{npsj})\left(
    \frac{\Oc_s \Un_n \Un_p}{\omega+i\eta-\epsilon_s - \epsilon_n - \epsilon_p}+
    \frac{\Oc_n \Oc_p \Un_s}{\omega-i\eta+\epsilon_p + \epsilon_n + \epsilon_s}\right)+\\
        +\sum_{n,p}(-i)\left(
    	\frac{\Oc_n \Un_p}{\epsilon_n + \epsilon_p} +
        	\frac{\Oc_p \Un_n}{\epsilon_p + \epsilon_n} \right)
         \left(v_{injp}-v_{inpj}\right) v^{m.f.}_{pn}
   \end{multline}
where $\Oc_i$ ($\Un_i$) is 1 if $i\in \Oc$ ($i\in \Un$) or 0 otherwise, $v_{ijkl}$ is defined by
\bea\label{vdef1}
\int d\vec{r}d\vec{r}'u^*_{i\sigma}(\vec{r})u^*_{l\lambda}(\vec{r}')
                v(|\vec{r}-\vec{r}'|)
                        u_{m\mu}(\vec{r}')u_{n\nu}(\vec{r})%&=&v_{i\sigma l\lambda m\mu n\nu}\\
                            &=&v_{ilmn}\delta_{\lambda \mu}\delta_{\sigma\nu}
\eea
for which $v_{jilm}=v_{ijml}$ and
\be\label{vdef2}
\int d\vec{r}u^*_{i\sigma}(\vec{r})
                v_{m.f.}(\vec{r}) u_{j\rho}(\vec{r})=v^{m.f.}_{i j}\delta_{\sigma\rho}
\ee
defines $v^{m.f.}_{ij}$.
%as explained in more detail in Appendix \ref{app:diagrams}.
By switching to the matrix notation
\be\label{matrixrep}
\{G_{ij}(\omega)\}\to {\bf G},\;\;\{g_{ij}(\omega)\}\to {\bf G_0},\;\;\{\Sigma^{(1)}_{ij}(\omega)\}\to {\bf \Sigma_1},\;\;\{\Sigma^{(2)}_{ij}(\omega)\}\to {\bf \Sigma_2}
\ee
one can readily apply formula \eqref{p11sigma} to calculate the finite version of \eqref{P11GF} in frequency domain.

We notice that the writing \eqref{P11GF} suggests that such an approximation can be cast 
in terms of \emph{two} Dyson equations.
Suppose we write
\be\label{twodyson}
\left\{\begin{array}{l}
	G(1,2)=G_0(1,2)+\int\!d3d4\; G_0(1,3)\Sigma(3,4) G(4,2)\\ \\
	\Sigma(1,2)=\Sigma_1(1,2)+\int\!d3d4\; \Sigma_1(1,3)\Xi(3,4) \Sigma(4,2)\\ 
\end{array}\right.
\ee
The approximant $P_{[1/1]}[G_0]$ is then obtained by approximating the kernel $\Xi$ with
\be\label{kernelapprox}
\Xi(1,2)\approx \int\!d3d4\;\Sigma_1^{-1}(1,3)\Sigma_2(3,4)\Sigma_1^{-1}(4,2).
\ee
More generally, the approximant $[N/N]$ can expressed as solution of a set of $N+1$
Dyson-like equations. For $N\to \infty$ the solution of such a hierarchy of Dyson equations
can be regarded as a continued fraction representation of the Green's function.

When solved by iteration, the second equation of \eqref{twodyson} with the approximation
\eqref{kernelapprox} gives rise to the approximate expansion for the self-energy:
\be
\Sigma\approx \Sigma_1+\Sigma_2\left(\Sigma_1^{-1}\Sigma_2+\Sigma_1^{-1}\Sigma_2\Sigma_1^{-1}\Sigma_2+\Sigma_1^{-1}\Sigma_2\Sigma_1^{-1}\Sigma_2\Sigma_1^{-1}\Sigma_2+...\right).
\ee
Although each term is of well defined order in $\lambda$, 
the formal parameter introduced to generate the perturbative expansion as in \eqref{matrixpert},
they do not correspond to specific diagrams.
More generally, kernels of the hierarchy of Dyson equations will always be combinations 
of terms of the perturbative expansion of the self-energy, but, 
while the latter can be expressed diagrammatically, 
the former cannot.
As we can see already in \eqref{kernelapprox}, 
this is due to the fact that, although products of diagrams are diagrams themselves,
not all products of diagrams and inverse of diagrams are 
always expressible as diagrams too.
The standard diagrammatic picture is therefore unsuitable 
to give a physical interpretation to such approximations.

Finally, we would like to comment about the computational cost of \eqref{P11GF}
and compare it with that of higher order approximants.
We observe that the calculation of the first nontrivial order
$P_{[1/1]}[G_0]$ for a real system requires: 
(\textrm{i}) the calculation of a truncated basis set,
(\textrm{ii}) the calculation of $v_{ijkl}$ given in \eqref{vdef1},
(\textrm{iii}) the calculation of $\Sigma^{(1)}_{ij}$ and $\Sigma^{(2)}_{ij}$ 
via formulas \eqref{sigmasimple1} and \eqref{sigmasimple2},
(\textrm{iv}) the calculation of \eqref{p11sigma} and then of $G$ using the Dyson equation,
or, equivalently, of \eqref{p11g}.
Usually the first step, if numerical, is performed by modern codes quite
efficiently even for large basis sets;
numerical computation and storage of $v_{ijkl}$, 
which involves a double integration on coordinate space
for each element of a four-index tensor, can, on the other hand, be quite demanding;
once that is available, computation and storage of $\Sigma^{(1)}_{ij}$ and $\Sigma^{(2)}_{ij}$ 
is less expensive, while the last step, which involves a matrix inversion, 
can also be an onerous task,
according to the size of the basis set required for convergence.
Now, suppose you want to calculate the next order $P_{[2/2]}[G_0]$.
Assuming that the basis set is already sufficiently large,
the cost of the second and forth points remains 
unaltered for the greatest part 
(the only difference being an undramatic extension of 
of \eqref{p11sigma} or \eqref{p11g}).
The main difference is calculating higher orders of the expansion of the self-energy,
which requires more diagrams that those contained in expressions 
\eqref{sigmasimple1} and \eqref{sigmasimple2}. Such a number increases exponentially,
which means that at certain point it will overcome the computational cost 
of (\textrm{ii}) and (\textrm{iv}). However, for sufficiently low orders,
this is still accessible (see for instance \cite{hirata2017} and references therein), 
making approximants like $P_{[2/2]}[G_0]$ probably still at hand.

%%%%%%%%%%%%%%%%%%%%%%%%%%%%%%%%%%%%%%%%%%%%%%%%%%%%%%%%%%%%%%%%%%%%%%%%%%%%%%%%%%%%%%%%%%
%%%%%%%%%%%%%%%%%%%%%%%%%%%%%%%%%%%%%%%%%%%%%%%%%%%%%%%%%%%%%%%%%%%%%%%%%%%%%%%%%%%%%%%%%%
%%%%%%%%%%%%%%%%%%%%%%%%%%%%%%%%%%%%%%%%%%%%%%%%%%%%%%%%%%%%%%%%%%%%%%%%%%%%%%%%%%%%%%%%%%

\section{Performance on Hubbard rings}\label{sec:comparison}

\subsection{The model}

The Hubbard model \cite{hubbard}, with its many incarnations and variants, is widely used in 
condensed matter physics for being a sufficiently simple model and 
yet reflecting many properties of real materials.
Its relevance for this work stems from the fact that in some simplified setups 
it can be exactly solved (at least numerically) while still representing a nontrivial many-body problem.
Here we consider $L$ number of contiguous sites filled with $N\leq 2L-2$ particles subject to the Hamiltonian:
\be\label{HHubbard}
H=-t\sum_{\langle i,j\rangle , \sigma} \hat{c}_{i\sigma}^\dagger \hat{c}_{j\sigma}
   +U\sum_{i}\hat{c}^\dagger_{i\uparrow}\hat{c}_{i\uparrow}
   			\hat{c}^\dagger_{i\downarrow}\hat{c}_{i\downarrow}
\ee
where $\hat{c}_{i,\sigma}^{(\dagger)}$ is an annihilation (creation) operator
of a fermion of spin $\sigma$ on a site $i$ and $\langle i,j\rangle$
denotes the sum over contiguous sites (two per site, in our one-dimensional setting), 
with periodic boundary conditions (which make it a so called `Hubbard ring').
Each site can accommodate up to two particles of opposite spin; particles can hop
to neighbouring sites with an energy gain of $-t$, while double occupancy of a site
is disfavoured by a repulsive on-site interaction U.
The zero-temperature Green's function of the model can be written as
\be\label{Ghubbard}
G_{i\sigma j\rho}(\omega):=
    \langle GS|\hat{c}_{i\sigma}\left(
    \omega+i \eta-\left(\hat{H}-\langle GS |\hat{H}|GS\rangle \right)\right)^{-1}
    \hat{c}^\dagger_{j\rho}|GS\rangle
    +\langle GS|\hat{c}^\dagger_{j\rho}\left(
    \omega-i \eta+\left(\hat{H}-\langle GS |\hat{H}|GS\rangle \right)\right)^{-1}
    \hat{c}_{i\sigma}|GS\rangle,
\ee
while the noninteracting one $G_0$ is defined as $\lim_{U\to 0}G$.

For sufficiently low number of sites $L$, the function \eqref{Ghubbard} %,
%or at least its diagonal part, 
can be calculated numerically.
In particular, we used a code \cite{alvarez2009,re:webDmrgPlusPlus} 
that relies on the Lanczos algorithm \cite{lanczos%
%,haydock}
}
for $L=4,6,8,10$ and $N=2,4,...,2L-2$.
The choice of having only states with an even number of particles,
which in their lowest energy configuration distribute in an unpolarized configuration
for which $G_{i\uparrow j\downarrow}(\omega)=G_{i\uparrow j\downarrow}(\omega)=0$
and $G_{i\uparrow j\uparrow}(\omega)=G_{i\downarrow j\downarrow}(\omega):= G_{i j}(\omega)$,
has the purpose of avoiding possible additional features due to spin polarization.

%%for $L\lesssim 12$ the diagonal part of \eqref{Ghubbard} 
%%can be easily calculated  using Haydock's method,
%%which relies on the Lanczos algorithm \cite{haydock}.
%%We use a code... \cite{alvarez2009,re:webDmrgPlusPlus} \walter{[some more explanation or just a good reference]}

%To avoid the rising of additional features due to spin polarization,
%we focused on states with equal number of particles of opposite spin, in which case we have 
%$G_{i\uparrow j\downarrow}(\omega)=G_{i\uparrow j\downarrow}(\omega)=0$
%and $G_{i\uparrow j\uparrow}(\omega)=G_{i\downarrow j\downarrow}(\omega):= G_{i j}(\omega)$.
%We are then able to calculate the exact spectral function
%\be
%A(\omega):=\frac{2}{\pi} \left| {\rm Im}\!\!\left[\sum_{i}G_{ii}(\omega)\right] \right|.
%\ee

\subsection{Goodness criterion}

\begin{figure*}[ht]
\includegraphics[scale=0.7]{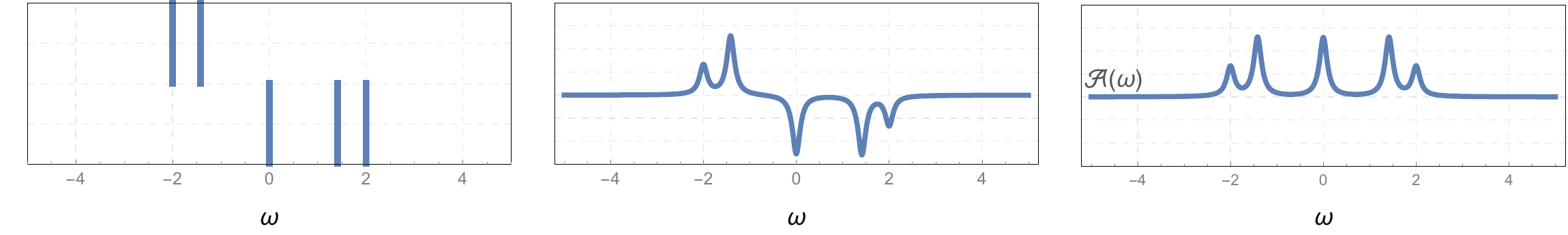}
\caption{
The spectral function of the Green's function is a collection of infinite peaks, here pictorially depicted in the first panel. When the $\eta$ parameter is considered finite, 
one has a sum of Lorenztian functions in the upper (`removal peaks') and the lower half plane (`addition peaks'), as in the second panel.
The smoothed spectral function, for which also addition
peaks lie on the upper half plane, is depicted in the last panel.
The absolute scale of these spectral functions is irrelevant,
as it depends on the unphysical $\eta$ parameter, 
and from here on values on the $y$-axis will always be omitted.
}
\label{finiteeta}
\end{figure*}

\begin{figure*}[ht]
\includegraphics[scale=0.7]{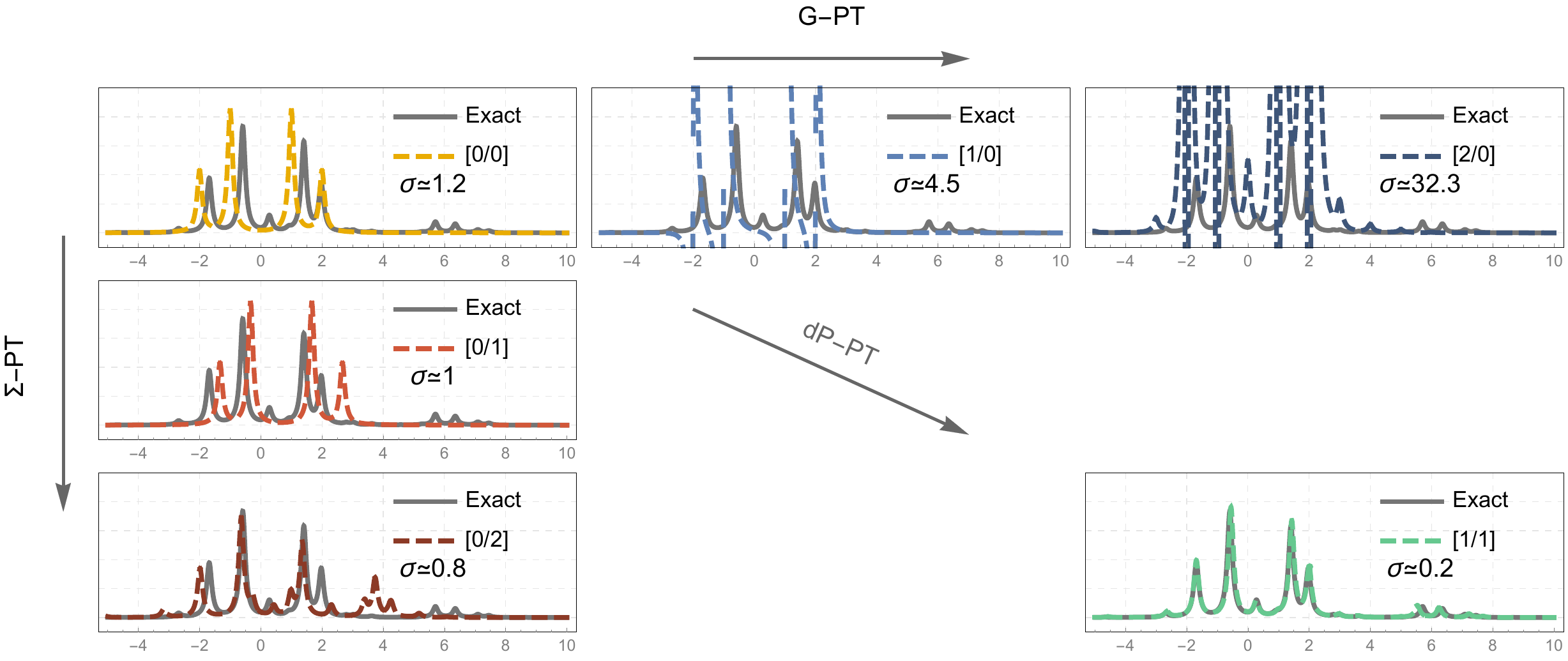}
\caption{The function $\mathpzc{A}(\omega)$ is here plotted
for different approximations in the case $L=6$, $N=2$, $U/t=4$.}
\label{fig:conv_case}
\end{figure*}

To establish the goodness of an approximation to $G$ we adopt a quantitative 
criterion based on the error in estimating the corresponding spectral function $A$, defined as:
\be
A(\omega):=\frac{2}{\pi} \left| {\rm Im}\!\!\left[\sum_{i}G_{ii}(\omega)\right] \right|.
\ee
Strictly speaking, the Green's function in the spectral representation, $G_{ij}(\omega)$,
is a distribution represented by a series of so called `addition' ($+$) and `removal' ($-$) poles
\be\label{spectral}
\frac{\alpha}{\omega-\omega_0 \pm i \eta}
\ee
with some real $\alpha$ and $\omega$ and an infinitesimal positive parameter $\eta$
that serves as reminder of the position of the pole in the complex plane when integrating
over the frequency $\omega$.
%If one takes the limit $\eta \to 0^+$ in \eqref{spectral},
%the result is a series of peaks of infinite hight and zero width.
In order to have  integrals over smooth  %, positive functions, first
functions we: {\rm i}) consider a finite value of $\eta$, for definiteness $\eta=0.1$,
which makes finite the height and width of the peak, and {\rm ii})
switch all poles to the upper part of the complex plane,
by setting $\alpha/(\omega-\omega_0 -i\eta)\to \alpha/(\omega-\omega_0 +i\eta)$,
which makes all peaks positive avoiding the derivative discontinuity that
we would have had if we had simply taken the absolute value.
This procedure, illustrated in figure \ref{finiteeta},
allows us to define the `smoothed' spectral function
\be\label{smoothedA}
\mathpzc{A}(\omega):= \frac{2}{\pi} \left. {\rm Im}\!\!\left[\sum_{i}G_{ii}(\omega)\right] \right|_{-i\eta\to +i\eta}
\ee
with $\eta=0.1$.
%%%%%%%%%%%%
%%%%%%%%%%%%
From that, we define the \emph{average absolute deviation} $\sigma$ as follow:
\be\label{sigma}
\sigma:= \frac{\int d\omega 
	\left| \mathpzc{A}_{exact}(\omega)-\mathpzc{A}_{approx}(\omega) \right|}{
		\int d\omega 
	\left| \mathpzc{A}_{exact}(\omega)\right|}.
\ee
This will be used as parameter to quantify the goodness of an approximation:
the higher is $\sigma$, the worst will be considered the corresponding approximation.
In figure \ref{fig:conv_case} we report a particularly neat example, the case with L=6, N=2, U/t=4
that illustrates the connection between the parameter $\sigma$ 
and the intuitive notion of `good approximate spectral function',
also providing a visual scale of reference for some representative values of $\sigma$.  
%%%%%%%%%%%%
%%%%%%%%%%%%
%%%%%%%%%%%%
%%%%%%%%%%%%
The criterion is obviously somehow arbitrary, for certain details of the spectral
function (position of the quasiparticle peaks, existence of satellites,...) 
may be of greater importance in certain situations; 
nonetheless it is general enough to provide an indication of the
behaviour of the three approaches under study under generic circumstances.

Concerning the calculation of the approximate spectral functions, two remarks are in order.
First, G-PT leads to approximate Green's functions that cannot be written 
in terms of poles\footnote{%
The underlying mechanism can be illustrated by:
\be
\frac{1}{\omega -\omega _0-i\eta-U}=
\frac{1}{\omega -\omega _0-i\eta}+\frac{U}{\left(\omega -\omega _0-i\eta\right)^2}
	+\frac{U^2}{\left(\omega -\omega _0-i\eta\right)^3}+...,
\ee
which, despite not being a sum of poles, still converges to the correct 
result for sufficiently small values of $U$.}, 
while $\Sigma$-PT and dP-PT always lead to approximations that can be reduced 
to sum of poles, even though we do not always do it in practice, 
the effects being negligible.
Second, in case of degeneracy, the expression \eqref{sigmasimple2} can be divergent,
but we found that %a simple procedure allows to carry out all calculations:
%in all degenerate cases considered it is sufficient to introduce a cutoff 
introducing a simple cutoff $\epsilon$ in formula \eqref{sigmasimple2} as
$(\epsilon^+-\epsilon^-)\to (\epsilon^+-\epsilon^- +\epsilon)$ 
and setting it to zero after the calculation of the Green's function
is sufficient to always get a finite result.%
%to proceed with 
%the algebraic manipulation that always leads to a perfect cancellation of the divergent 
%terms and a result that is $\lambda$-independent.
\footnote{%
Unravelling the intricacies of perturbation theory on degenerate states 
lies beyond the purposes of this work,
the interested reader being referred to \cite{brouder2011}.
}

\subsection{Choice of the basis}

The Green's function of the Hubbard model \eqref{Ghubbard}
has been here defined in the so called `site' basis.
Any Bogolioubov transformation of the ladder operators
induces a new basis, which results in a rotation of the matrix $\{G_{ij}(\omega)\}$.
The choice of the basis does not affect the spectral function,
which is calculated from the trace of the Green's function.
However, \pade approximants built in different basis are inequivalent.
This rises the question:
is there a basis in which \pade approximants 
(not necessarily diagonal ones, but also G-PT and $\Sigma$-PT) work better?
To attempt to reply to this question, we looked at the case of $L=2$, 
also known as Hubbard `dimer'.

For two sites the model is simple enough to admit an analytic solution
and in Appendix \ref{app:dimer} we show the groundstates,
labelled by the number of particles $\langle \hat{N} \rangle$ and, in case of degeneracy,
the spin polarization $\langle \hat{S}_z \rangle$.
\pade approximants can be calculated directly from the expression of $G$
one obtains simplifying \eqref{Ghubbard} with those groundstates rather than using 
many-body perturbation theory \eqref{pert}.
This allows to easily calculate approximants of very high order.
Moreover,  approximants in different basis can be calculated by considering rotations of $G$.
A simple Bogolioubov transformation, 
reported in Appendix \ref{app:dimer}, is sufficient to make $G$ diagonal,
independently of the number of particles and interaction strength.
The fact that such a transformation does not depend on $U$
implies that also $G_0$ is diagonal in this basis. 
%In fact, $G_0$ and $G$ are diagonalized by the same 
%$U-$independent rotation for arbitrary number of sites, not only $L=2$. \walter{check}

Such a specific basis is particularly relevant because we found that, in the case of the dimer,
while G-PT is not affected by a change of basis,
both $\Sigma$-PT and dP-PT work better in this basis than any other.
In fact, dP-PT converged to the exact result in only one step,
namely $P_{[1/1]}[G_0]=G$, for $L=2$, $N=0,1,2,3,4$,
as shown in Appendix \ref{app:dimer} for the representative case $N=2$.
Such a remarkable property of this basis for the case $L=2$ 
suggested us to formulate the following conjecture:
\pade approximants work best in the basis in which the noninteracting 
Green's function is diagonal, irrespective of the number of sites or particles.
From now on, we shall assume that and all calculations for $L>2$ will be shown
only in the basis in which $G_0$ is diagonal.

\subsection{Approximations comparison}

Combining formulae
(\ref{p11sigma},\ref{sigmasimple1},\ref{sigmasimple2},\ref{matrixrep}) with $v^{m.f.}_{ij}=0$
we have an explicit expression
for the functional $P_{[1/1]}$  
that also applies to the Hubbard model.
Such an approximant is made with the same amount of information 
(i.e. same diagrams) contained in the second order expansion of $\Sigma$ or $G$,
denoted by $[0/2]$ and $[2/0]$ in the \pade notation.
To compare the behaviour of the different expansions 
G-PT, $\Sigma$-PT and dP-PT we shall then consider the following sequences:
\begin{itemize}
    \item $([0/0],[1/0],[2/0])$ for G-PT,
    \item $([0/0],[0/1],[0/2])$ for $\Sigma$-PT,
    \item and $([0/0],[1/1])$ for dP-PT.
\end{itemize}

\begin{figure}[ht]
    \includegraphics[scale=1.2]{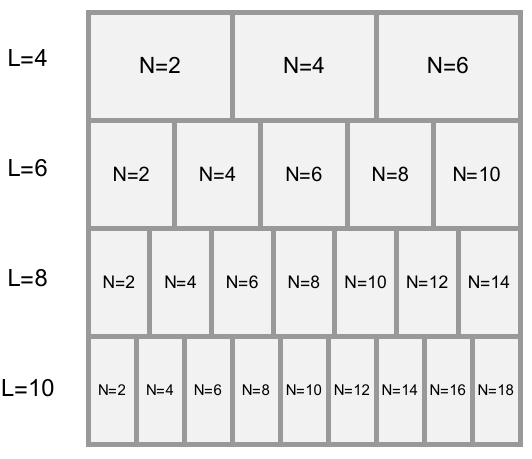}
    \caption{}
    \label{legend}
\end{figure}

\begin{figure*}[ht]
    \includegraphics[scale=0.7]{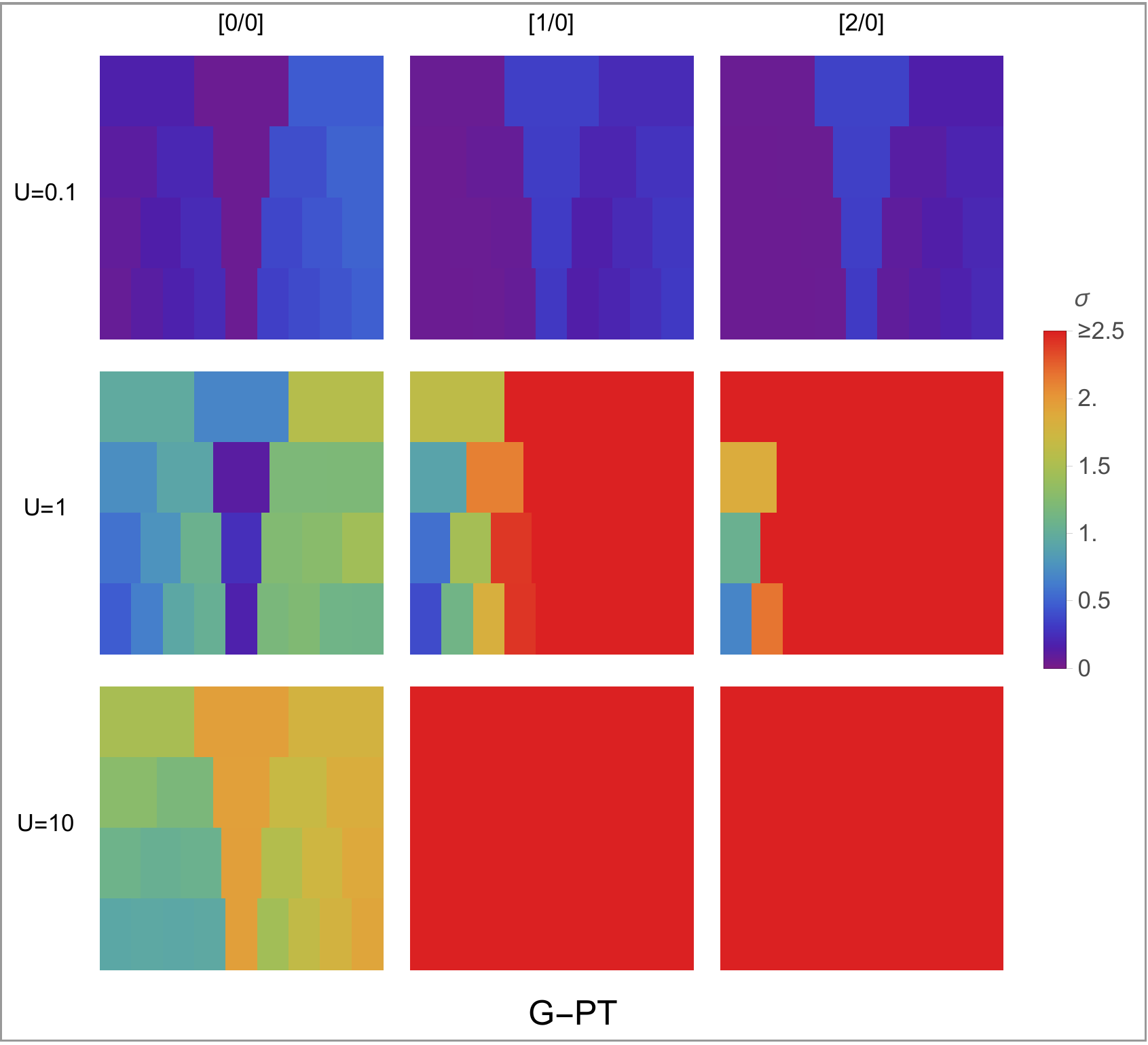}
    \caption{}
    \label{conv_PN0}
\end{figure*}

We have calculated the deviation $\sigma$, as defined in \eqref{sigma},
for the spectral functions arising from the corresponding approximations 
to the Green's function, for all rings ($L=4,6,8,10$) and fillings ($N=2,4,...,2L-2$),
in three different regimes of interaction:
a weak, $U=0.1$, an intermediate, $U=1$, and a strong coupling, $U=10$, in units of $t$.
For a given approximation $[N/M]$ and interaction strength $U$,
the values of $\sigma$ for different sites and particles are here plotted on 
a single panel, arranged as reported in legend \ref{legend}.
This allows to get an idea of a specific approximation for a given interaction strength
for all systems considered at a glance.
Panels are then grouped for sequence of approximations, G-PT in fig. \ref{conv_PN0},
 $\Sigma$-PT in fig. \ref{conv_P0N} and dP-PT in fig. \ref{conv_PNN}.
Numerical values are explicitly reported in Appendix \ref{app:data}.

\begin{figure*}[ht]
    \includegraphics[scale=0.7]{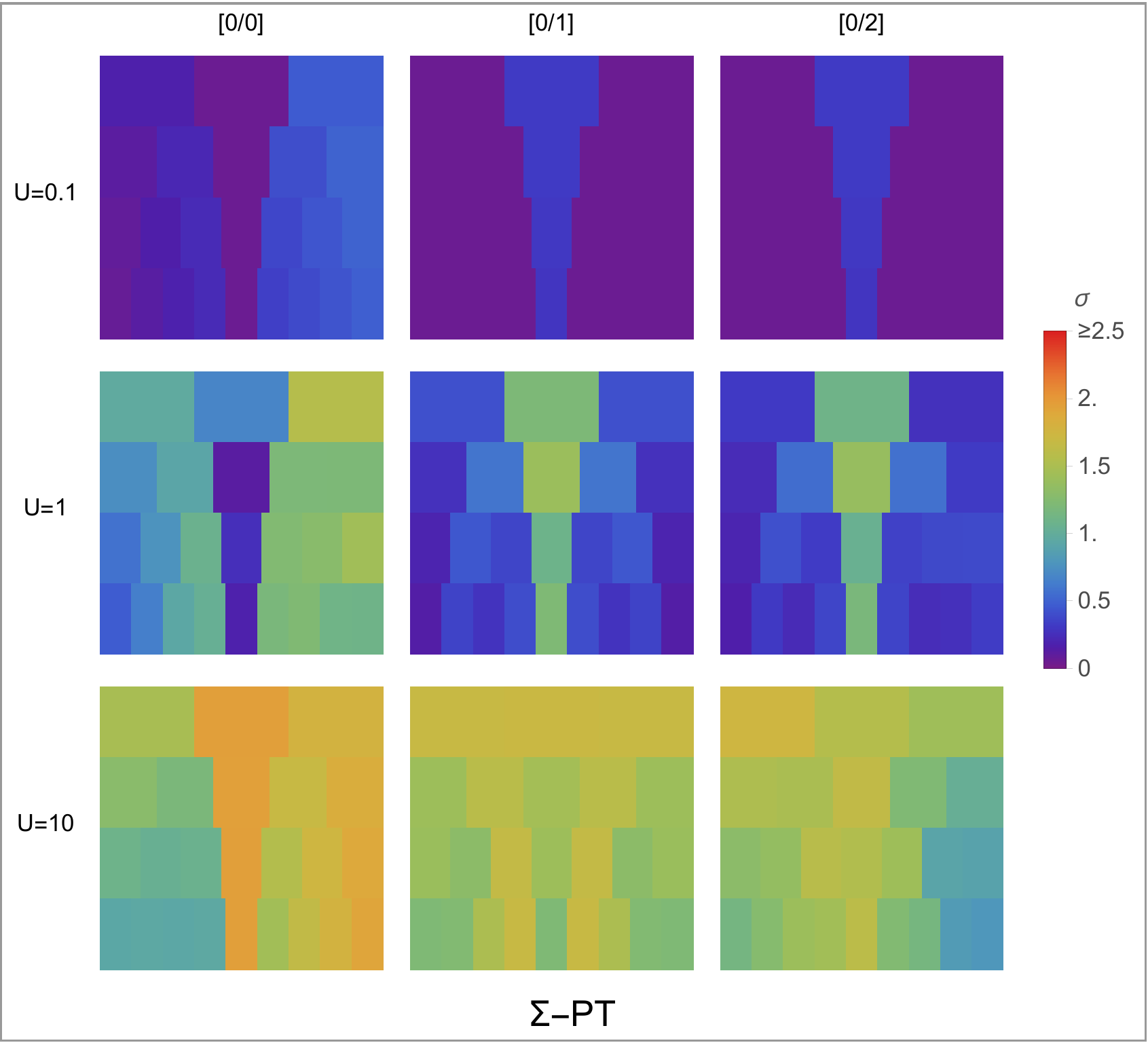}
    \caption{}
    \label{conv_P0N}
\end{figure*}
\begin{figure}[ht]
    \includegraphics[scale=0.7]{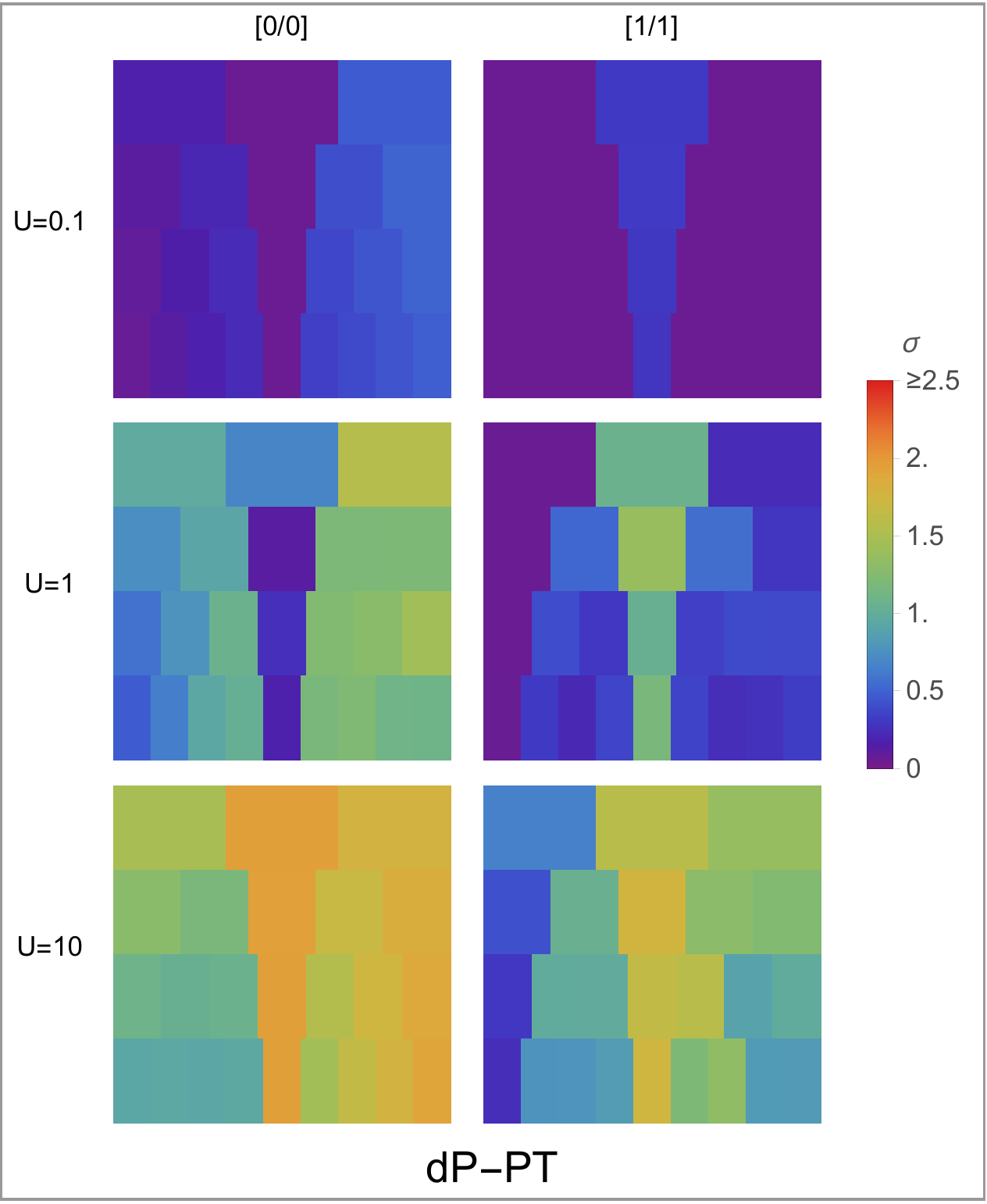}
    \caption{}
    \label{conv_PNN}
\end{figure}
\begin{figure*}[ht]
\includegraphics[scale=0.5]{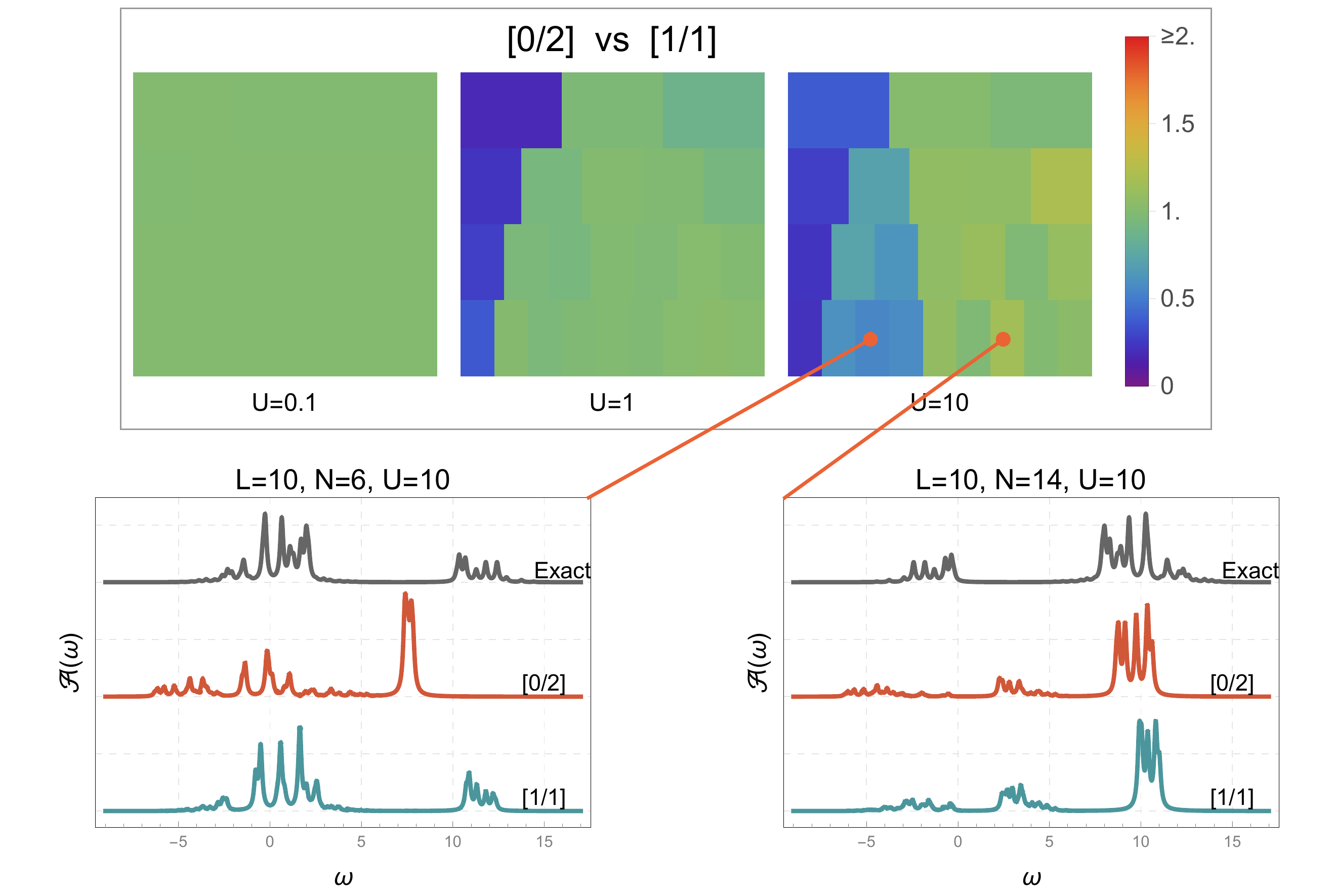}
\caption{\textit{Top panel} The ratio $\sigma_{[1/1]}/\sigma_{[0/2]}$ is plotted.
For most cases this is less than one (green to blue color),
indicating that $[1/1]$ provides a better estimate
of the exact function than $[0/2]$.
\textit{Bottom panel} The spectral functions $\mathpzc{A}(\omega)$ 
in two cases representative of the situations
$\sigma_{[1/1]}/\sigma_{[0/2]}<1$ and $\sigma_{[1/1]}/\sigma_{[0/2]}>1$
are plotted, 
for a better visual reference of the approximations.
}
\label{fig:P02vP11}
\end{figure*}

Recalling that the closer to 0 (blue in plots) the better is the approximation,
we recognize the following trends:
\begin{itemize}
	\item all approximations deteriorate with increasing $U$;
    \item varying the number of sites $L$ does not seem to change the 
    quality of the approximations; a higher number of sites 
    only seems to increase the `definition' of the plots;
    \item almost all approximations deteriorates when approaching the half-filling case;
    \item G-PT leads to meaningful results only 
    	in the weakly coupling regime, while $\Sigma$-PT and dP-PT
        lead to sensible approximations also in the intermediate and 
        strong one;
    \item $\Sigma$-PT seems to have a slow convergence rate;
    namely the term $[0/2]$ does not improve over $[0/1]$  as much as $[0/1]$ does over $[0/0]$;
    \item dP-PT seems to have a higher rate of convergence (even when compared with the sequence $([0/0],[0/2])$),
    especially in the low filling cases ($N<L$);
    \item the higher the $U$ the better is dP-PT over $\Sigma$-PT.
\end{itemize}
A more direct comparison of $\Sigma$-PT and dP-PT is reported in figure \ref{fig:P02vP11}, top panel.
As anticipated, $\Sigma$-PT is generally worst than dP-PT, especially for low fillings and high values of $U$.
The cases in which $\Sigma$-PT is better are only a few and the improvement 
is not as much as that of dP-PT in other cases,
as exemplified in the bottom panels of figure \ref{fig:P02vP11}.

\begin{figure}[ht]
	\includegraphics[scale=0.7]{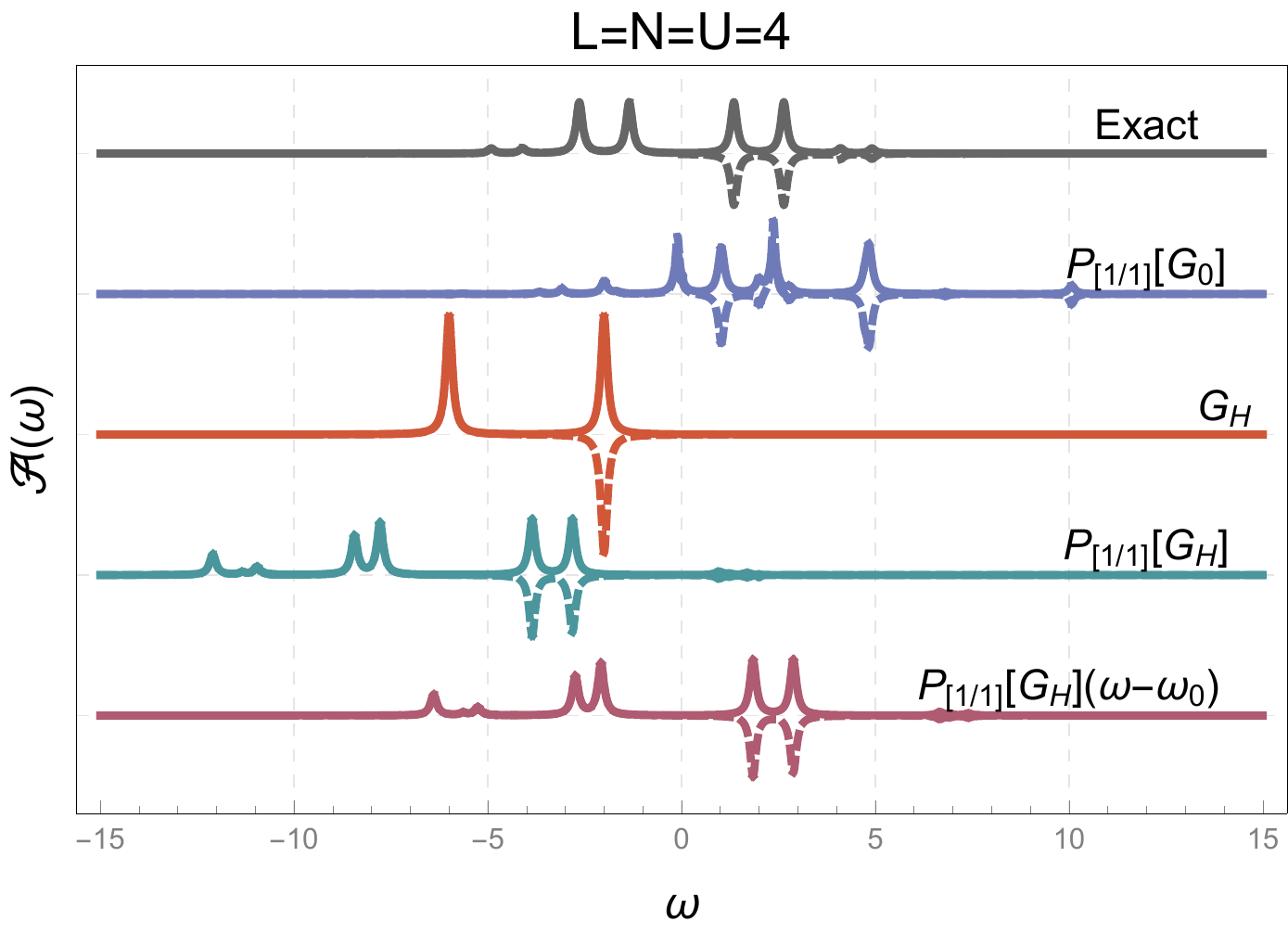}
	\caption{Various approximated spectral functions
	are compared to the exact one in the case $L=N=U/t=4$.
	To distinguish addition from removal peaks,
	the former are plotted in dashed lines in the lower half of the plane.
	}
	\label{fig:P11GH}
\end{figure}

In absolute terms, $[1/1]$ leads to 
a decent approximation in most cases 
(as seen in figures \ref{fig:conv_case} and \ref{fig:P02vP11}),
but, like all other approximations here considered,
has serious problems in capturing the half-filling case,
as one can see by comparing the first two top curves of
figure \ref{fig:P11GH}.
Even though the purpose of our study
is to compare dP-PT to other perturbative approaches
and not necessarily to provide precise estimates in absolute terms, we would like to focus a bit more 
%on the half-filling case which bares special physical relevance.
%For finite temperature and infinite number of sites, the half-filling case
%is known to have interesting physical properties, such as Mott insulating behavior,
%anti-ferromagnetic order, etc. \cite{scalettar},
%and it is known to represents perhaps the most challenging case to capture with standard perturbative approximations.
%For our zero-temperature Hubbard rings, the half-filled case already shows some signs
%of the interesting physics displayed by more complicated setups, like, 
%for instance, for $L=N=4$,
%the gap (energy difference between first addition and last removal peaks)
%is zero for $U=0$ but positive for $U>0$. 
%%% NEW (start) %%%
on the half-filling case which in general (i.e. not just rings) 
bares special physical relevance \cite{scalettar}.
For instance, for $L=N=4$,
the gap, defined as the energy difference between first addition and last removal peaks,
is zero for $U=0$ but positive for $U>0$,
which can be regarded as a sign of strong correlation. %
%Mott transition, typical of strongly correlated systems.
%%% NEW (end) %%%
In such a case, shown in figure \ref{fig:P11GH}
 for $U/t=4$, the approximant
 $P_{[1/1]}[G_0]$  does not capture the (anti)symmetry of the exact spectrum, 
nor the fact that addition and removal peaks are well separated by the gap,
leading to a spectral function that one may deem as quite far from the exact one.
It is legitimate to expect that higher order dP-PT approximants correct these problems and 
converge towards the exact function, even though 
%we have no solid elements to believe that that would happen, and
%with a reasonable number of terms.
we have no elements to actually prove that and, even then, that convergence
is practically achievable. 
On the other hand, for all half-filled systems here considered, we did notice that a sensible improvement 
comes already from considering a starting point for $P_{[1/1]}$ different from the pure noninteracting Green's function.
By taking $G_H$, defined as solution to the equation
\be
G_H=G_0+G_0\Sigma_H[G_H]G_H
\ee
with  
\be
\Sigma^{(H)}_{ij}[g]:=
-\sum_{kl}\Int \frac{d\omega'}{2\pi} 
i g_{kl}(\omega')v_{ilkj}
\ee
and building $P_{[1/1]}[G_H]$ using formulae (\ref{p11sigma},\ref{sigmasimple1},\ref{sigmasimple2},\ref{matrixrep}) with $v^{m.f.}_{ij}=v^{\rm H}_{ij}=-i\delta_{ij}\Sigma^{(H)}_{ij}[g]$,
one gets to a spectral function that partially restores the (anti)symmetry of the spectral function
and correctly opens a gap between addition and removal peaks,
as one can see in the penultimate (from top) curve of figure \ref{fig:P11GH}. 
Finally, an overall shift with no physical significance can be tuned 
to give rise to what we would deem as quite a decent approximation to the exact function. 
By comparing $P_{[1/1]}[G_H]$ to $P_{[0/0]}[G_H]$, which is simply $G_H$ (third curved in the plot), 
it seems then reasonable to expect that of $P_{[n/n]}[G_H]$ for $n>1$ 
would provide even better approximations to the exact $G$.

\section{Conclusions}

\pade approximants have proved to be an effective tool to get estimates 
of unknown functions in many fields of physics an mathematics 
\cite{bakergraves,baker1971}.
Their use in the calculation of electronic properties has been only partially explored,
and in particular in the calculation of Green's functions it has been limited so far 
as ancillary tool, like for the extrapolation of the analytic continuation 
of the Matsubara Green's functions,
or to model systems (see, for instance, 
\cite{goodson2012,schott2016,pavlyukh2017} 
and references therein.)

We have here presented a way to build \pade approximants of Green's functions
that generalizes the standard approximations based on
the perturbative expansion of $G$  or the self-energy $\Sigma$ and provides
a general framework suitable for model as well as for real systems.
Moreover, we put forth the conjecture that, 
among all possible \pade approximants $[p/q]$
of given order $p+q$ of the Green's function,
diagonal ones, $p=q$, offer the best approximation.
As preliminary test, we compared diagonal \pade approximants of order $[1/1]$ 
against approximations arising from direct perturbative expansion of $G$ 
and perturbative expansion of the self-energy of equivalent order
for a series of solvable models, namely Hubbard rings with various number of sites,
fillings and interaction strength,
whose exact Green's functions are (numerically) known.
%Based on a measure of likeness of spectral functions, we found indeed that
%diagonal \pade approximants offered the best approximation  in the vast majority of the cases,
%for all other still being relatively good.
Based on a measure of likeness of spectral functions, 
we found indeed that in general the diagonal \pade approximant offered 
the most reliable approximation.
In the great majority of cases it overcomes the other approximations,
for all remaining cases still being quite close to the best (second-order $\Sigma$-PT).
Particularly good results were obtained for high values of the interaction $U\gg t$
and low fillings $N<L$, irrespectively of the number $L$ of sites.
We also presented a case of physical relevance ($L=N=U/t=4$)
in which the approximant $[1/1]$ built on a mean field, 
rather than completely noninteracting, Green's function
greatly improves the otherwise not so good, in absolute terms, approximation.
%We are then led to conclude that diagonal \pade approximants
%may offer an important tool in estimating Green's functions of real systems.
%In particular, already the first nontrivial approximant $P_{[1/1]}[G_0]$
%may reveal to provide a sufficiently good estimate for the Green's function 
%of real systems, based on the success of methods like GF2 and `one-shot' GWA,
%whose....

Diagonal \pade approximants of the Green's function 
%This approximation is not based on physical ground.
were not directly built on some physical principle 
and in fact their physical interpretation 
as a resummation of certain terms of the 
pertubative series needs further investigation.
On one hand this might make uncomfortable who would 
try to anticipate its behavior on a specific system.
On the other hand one can say, in a rather optimistic attitude,
that there is great potential in an approximation that is not designed 
around specific features of a class of systems.
The performance of $[1/1]$ on the model studied in this work 
is a first prove of that.

A more tangible advantage of diagonal \pade approximants $[N/N]$ 
is that they are systematically improvable by increasing the order $N$.
In this regard, we also argued that the computational cost increase
implied by the rising of the order $N$ would probably remain 
moderate for a few orders.
This is of capital importance in view of having a reliable, predictive tool.

In fact, the spreading of approximations based on the second-order expansion 
of the self-energy for a wide range of applications 
(from nuclear \cite{soma1} to molecular \cite{phillips2014} to periodic systems \cite{rusakov2016})
suggests that already $P_{[1/1]}$, which in our study improves on $[0/2]$ 
almost in all cases, may already provide a competitive approximation,
in all those situations and beyond.\footnote{
It must be noticed that current approaches based on 
the second-order expansion of the self-energy are generally 
formulated within a self-consistent formalism, according to which
diagrams of the expansion of the self-energy are not written in terms 
of $G_0$ but of the fully interacting $G$ itself.
While this alters the diagrammatic content of the generic term
$\Sigma_n$, formula \eqref{P11SIGMA} remains valid and the right-hand side
can be read as a functional of $G$ if $\Sigma_1$ and $\Sigma_2$ are, too.}

In conclusions, we believe that diagonal \pade approximants certainly deserve
more attention in the study of the many-body problem,
as they may provide an effective new route for designing reliable, 
systematic approximations within the framework of standard perturbation theory.

\appendix
\section{}\label{app:diagrams}

When projected on a basis and Fourier-transformed to frequency space,
diagrams \eqref{diagrams} can be written as:
\be\label{Sigmafirstorder}
\Sigma^{(1)}_{ij}\equiv\includegraphics[scale=0.5]{diaga}+\includegraphics[scale=0.5]{diagb}+\includegraphics[scale=0.8]{diagc}=
\sum_{kl}\Int \frac{d\omega'}{2\pi} 
i g_{kl}(\omega')\left( v_{iljk}- 
	v_{ilkj}\right)-i\delta_{ij}v^{m.f.}_{ij}
\ee
and
\be
\Sigma^{(2)}_{ij}(\omega)\equiv 
\Sigma^{(2.1)}_{ij}+\Sigma^{(2.2)}_{ij}
+\Sigma^{(2.3)}_{ij}+\Sigma^{(2.4)}_{ij}(\omega)
+\Sigma^{(2.5)}_{ij}+\Sigma^{(2.6)}_{ij}(\omega)
+\Sigma^{(2.7)}_{ij}+\Sigma^{(2.8)}_{ij}
\ee
\bea
\Sigma^{(2.1)}_{ij}&\equiv& \includegraphics[scale=0.8]{diag1}=
-\sum_{nopqrs}\Int \frac{d\omega'd\omega''}{4\pi^2} 
	g_{no}(\omega')g_{pq}(\omega'')g_{rs}(\omega')v_{iorj}v_{qsnp}
\eea
\bea
\Sigma^{(2.2)}_{ij}&\equiv& \includegraphics[scale=0.8]{diag2}=
\sum_{nopqrs}\Int \frac{d\omega'd\omega''}{4\pi^2}g_{no}(\omega')g_{pq}(\omega'')g_{rs}(\omega')v_{iojr}v_{qsnp}
\eea
\bea
\Sigma^{(2.3)}_{ij}&\equiv& \includegraphics[scale=0.8]{diag3}=
\sum_{nopqrs}\Int \frac{d\omega'd\omega''}{4\pi^2} 
	g_{no}(\omega')g_{pq}(\omega'')g_{rs}(\omega'')v_{iqrj}v_{osnp}
\eea
\bea
\Sigma^{(2.4)}_{ij}(\omega)&\equiv& \includegraphics[scale=0.8]{diag4}=
-\sum_{nopqrs}\Int \frac{d\omega'd\omega''}{4\pi^2}g_{no}(\omega+\omega'-\omega'')g_{pq}(\omega'')g_{rs}(\omega')v_{isnp}v_{oqrj}
\eea
\bea
\Sigma^{(2.5)}_{ij}&\equiv& \includegraphics[scale=0.8]{diag5}=
-\sum_{nopqrs}\Int \frac{d\omega'd\omega''}{4\pi^2} g_{no}(\omega')g_{pq}(\omega'')g_{rs}(\omega'')v_{iqjr}v_{osnp}
\eea
\bea
\Sigma^{(2.6)}_{ij}(\omega)&\equiv& \includegraphics[scale=0.8]{diag6}=
\sum_{nopqrs}\Int \frac{d\omega'd\omega''}{4\pi^2} g_{no}(\omega+\omega'-\omega'')g_{pq}(\omega'')g_{rs}(\omega')v_{isnp}v_{oqjr} 
\eea
\bea
\Sigma^{(2.7)}_{ij}&\equiv&
%\begin{feynman}
%    \electroweak[color=000000, endcaps=true, flip=false, labelDistance=0.42, labelLocation=0.50, showArrow=false, lineWidth=1.4285714285714286]{0, 0}{1, 0}
%    \fermion[color=000000, endcaps=true, flip=false, labelDistance=0.42, labelLocation=0.50, showArrow=true, lineWidth=1.4285714285714286]{0, 0}{0, 1}
%    \fermion[color=000000, endcaps=true, flip=false, labelDistance=0.42, labelLocation=0.50, showArrow=true, lineWidth=1.4285714285714286]{1, 1}{1, 0}
%    \dashed[color=000000, endcaps=true, flip=false, labelDistance=0.42, labelLocation=0.50, showArrow=false, lineWidth=1.4285714285714286]{0, 1}{1, 1}
%\end{feynman}=
\includegraphics[scale=0.6]{diag7}=
\sum_{nopq}\Int \frac{d\omega'}{2\pi}
g_{no}(\omega')g_{pq}(\omega')v_{iojp}v^{m.f.}_{qn}
\eea
\bea
\Sigma^{(2.8)}_{ij}&\equiv& \includegraphics[scale=0.6]{diag8}=
-\sum_{nopq}\Int \frac{d\omega'}{2\pi} 
	g_{no}(\omega')g_{pq}(\omega')v_{iopj}v^{m.f.}_{qn}
\eea
where $g_{ij}(\omega)$ is the (Fourier-transformed $ij$-component of) noninteracting Green's function, $v_{iojp}$ and $v^{m.f.}_{ij}$
are defined via \eqref{vdef1}, \eqref{vdef2}, and the integral is taken
over a closed anticlockwise contour large enough to include 
all poles of the integrand on the upper-half of the complex plane.
When $g_{ij}(\omega)$ is written as in \eqref{gij}
those expressions simplify to \eqref{sigmasimple1} and \eqref{sigmasimple2}
by means of the following identities:
\be\label{intg}
\Int \frac{d\omega}{2\pi i}g_{jj}(\omega)=\left\{
\begin{array}{lr}
1 & j\in\Oc\\
0 & j\in\Un
\end{array}
\right.
\ee
\be\label{intgg}
\Int \frac{d\omega}{2\pi i}g_{pp}(\omega)g_{qq}(\omega)=\left\{
\begin{array}{lr}
-\frac{1}{\epsilon^-_p + \epsilon^+_q} & \begin{array}{l} p\in\Oc\\q\in\Un\end{array}\\
&\\
-\frac{1}{\epsilon^-_q + \epsilon^+_p} & \begin{array}{l} q\in\Oc\\p\in\Un\end{array}\\
&\\
0 & otherwise
\end{array}
\right.
\ee
and
\be\label{intgg2}
\Int \frac{d\omega'}{2\pi i}g_{pp}(\omega'-\omega)g_{qq}(\omega')=\left\{
\begin{array}{lr}
-\frac{1}{\omega-i\eta+\epsilon^-_q + \epsilon^+_p} & \begin{array}{l} p\in\Un\\q\in\Oc\end{array}\\
&\\
\frac{1}{\omega+i\eta-\epsilon^-_p - \epsilon^+_q} & \begin{array}{l} q\in\Un\\p\in\Oc\end{array}\\
&\\
0 & otherwise
\end{array}
\right.
\ee
which comes from
\begin{multline}
\Int \frac{d\omega'}{2\pi i}\left(
	\frac{\alpha_1^+}{\omega'-\omega+i\eta+\omega_1^+} 
		+ \frac{\alpha_1^-}{\omega'-\omega-i\eta+\omega_1^-}
	\right)\left(
	\frac{\alpha_2^+}{\omega'+i\eta+\omega_2^+} 
		+ \frac{\alpha_2^-}{\omega'-i\eta+\omega_2^-}\right)=\\
=\frac{\alpha_1^- \alpha_2^+}{\omega+i\eta-\omega_1^- + \omega_2^+}
	-\frac{\alpha_1^+ \alpha_2^-}{\omega-i\eta+\omega_2^--\omega^+_1}.
\end{multline}

\section{}\label{app:dimer}
In the case $L=2$ the Hamiltonian \eqref{HHubbard} reduces to 
\be
\hat{H}=
-t\sum_{\sigma=\uparrow,\downarrow}\left(
\hat{c}^\dagger_{1\sigma}\hat{c}_{2\sigma}+\hat{c}^\dagger_{2\sigma}\hat{c}_{1\sigma}
\right)
+U\sum_{i=1,2}\left(
\hat{c}^\dagger_{i\downarrow}\hat{c}_{i\downarrow}\hat{c}^\dagger_{i\uparrow}\hat{c}_{i\uparrow}
\right).
\ee
whose groundstates are reported in Table \ref{dimerGS}.

\begin{table*}[ht]
\centering
\begin{tabular}{ |c|c|c|c| } 
%\hline
\toprule
$\langle \hat{N}\rangle$ & $\langle \hat{S}_z\rangle$&$\langle \hat{H}\rangle$ & $|GS\rangle$  \\  \hline
0 & 0 & 0 & $| 0\rangle$ \\
1 & $\frac{1}{2} \cos(2 \phi)$ & $-t$  & $\frac{1}{\sqrt{2}}\left(
\cos(\phi)\left(\hat{c}^\dagger_{1\uparrow}+\hat{c}^\dagger_{2\uparrow}\right)+
\sin(\phi)\left(\hat{c}^\dagger_{1\downarrow}+\hat{c}^\dagger_{2\downarrow}\right)
\right) | 0\rangle$ \\
2 & 0 & $\frac{1}{2} \left(U-\sqrt{16 t^2+U^2}\right)$ & $ 
\frac{1}{2}\left(
\sqrt{1-\frac{U}{\sqrt{16 t^2+U^2}}}\left(
\hat{c}^\dagger_{1\uparrow}\hat{c}^\dagger_{1\downarrow}
+\hat{c}^\dagger_{2\uparrow}\hat{c}^\dagger_{2\downarrow}
\right)
+
\sqrt{1+\frac{U}{\sqrt{16 t^2+U^2}}}\left(
\hat{c}^\dagger_{1\uparrow}\hat{c}^\dagger_{2\downarrow}
+\hat{c}^\dagger_{2\uparrow}\hat{c}^\dagger_{1\downarrow}
\right)
\right)| 0\rangle$ \\
3 & $\frac{1}{2} \cos(2 \phi)$ & $-t+U$ & $ \frac{1}{2}\left(
\cos(\phi)\left(
\hat{c}^\dagger_{1\uparrow}\hat{c}^\dagger_{2\uparrow}\hat{c}^\dagger_{2\downarrow}
-\hat{c}^\dagger_{1\uparrow}\hat{c}^\dagger_{1\downarrow}\hat{c}^\dagger_{2\uparrow}
\right)+
\sin(\phi)\left(
\hat{c}^\dagger_{1\downarrow}\hat{c}^\dagger_{2\uparrow}\hat{c}^\dagger_{2\downarrow}
-\hat{c}^\dagger_{1\uparrow}\hat{c}^\dagger_{1\downarrow}\hat{c}^\dagger_{2\downarrow}
\right)
\right)| 0\rangle$ \\
4 & 0 & $2U$ & $ 
\hat{c}^\dagger_{1\uparrow}\hat{c}^\dagger_{1\downarrow}\hat{c}^\dagger_{2\uparrow}\hat{c}^\dagger_{2\downarrow}| 0\rangle$ \\% \bottomrule
\hline
\end{tabular}
\caption{}\label{dimerGS}
\end{table*}
From those one can calculate the Green's function from the definition
\be
G_{i j\sigma\rho}(\omega)=\langle
\hat{c}_{i\sigma}\left(
\omega+i\eta-\left(\hat{H}-\langle \hat{H}\rangle\right)
\right)^{-1}\hat{c}^\dagger_{j\rho}\rangle
+\langle\hat{c}^\dagger_{j\rho}\left(
\omega-i\eta+\left(\hat{H}-\langle \hat{H}\rangle\right)
\right)^{-1}\hat{c}_{i\sigma}
\rangle.
\ee
If this is performed in the basis $i,j=a,b$,
and $\sigma,\rho=h,g$, defined by the transformation
\be\label{hellogoodbye}
\begin{array}{c}
	\hat{c}_{1\uparrow} =
	\frac{\left(\cos \left(\frac{\phi }{2}\right)+\sin \left(\frac{\phi }{2}\right)\right) \left(\cos \left(\frac{\phi }{2}\right) \left(\hat{c}_{a h}+\hat{c}_{b
			h}\right)-\sin \left(\frac{\phi }{2}\right) \left(\hat{c}_{a g}+\hat{c}_{b g}\right)\right)}{\sqrt{2} \sqrt{\sin (\phi )+1}}
	\\
	\hat{c}_{1\downarrow}=
	\frac{\left(\cos \left(\frac{\phi }{2}\right)+\sin \left(\frac{\phi }{2}\right)\right) \left(\cos \left(\frac{\phi }{2}\right) \left(\hat{c}_{a g}+\hat{c}_{b
			g}\right)+\sin \left(\frac{\phi }{2}\right) \left(\hat{c}_{a h}+\hat{c}_{b h}\right)\right)}{\sqrt{2} \sqrt{\sin (\phi )+1}}
	\\
	\hat{c}_{2 \uparrow}=
	-\frac{\left(\cos \left(\frac{\phi }{2}\right)+\sin \left(\frac{\phi }{2}\right)\right) \left(\sin \left(\frac{\phi }{2}\right) \left(\hat{c}_{b g}-\hat{c}_{a
			g}\right)+\cos \left(\frac{\phi }{2}\right) \left(\hat{c}_{a h}-\hat{c}_{b h}\right)\right)}{\sqrt{2} \sqrt{\sin (\phi )+1}} 
	\\
	\hat{c}_{2 \downarrow}=
	-\frac{\left(\cos \left(\frac{\phi }{2}\right)+\sin \left(\frac{\phi }{2}\right)\right) \left(\cos \left(\frac{\phi }{2}\right) \left(\hat{c}_{a g}-\hat{c}_{b
			g}\right)+\sin \left(\frac{\phi }{2}\right) \left(\hat{c}_{a h}-\hat{c}_{b h}\right)\right)}{\sqrt{2} \sqrt{\sin (\phi )+1}}
\end{array}
\ee
then $G_{ij\sigma\rho}=\delta_{ij}\delta_{\sigma\rho}G_{ij\sigma\rho}$. In particular, for $N=2$ one has
\be
G_{aahh}=\frac{\frac{1}{2}-\frac{2 t}{\sqrt{16 t^2+U^2}}}{\frac{1}{2} \left(2
   t-U+\sqrt{16 t^2+U^2}\right)-i \eta +\omega }+\frac{\frac{2 t}{\sqrt{16
   t^2+U^2}}+\frac{1}{2}}{\frac{1}{2} \left(2 t-U-\sqrt{16
   t^2+U^2}\right)+i \eta +\omega }
\ee
\be
G_{bbhh}=\frac{\frac{1}{2}-\frac{2 t}{\sqrt{16 t^2+U^2}}}{\frac{1}{2} \left(-2
   t-U-\sqrt{16 t^2+U^2}\right)+i \eta +\omega }+\frac{\frac{2 t}{\sqrt{16
   t^2+U^2}}+\frac{1}{2}}{\frac{1}{2} \left(-2 t-U+\sqrt{16
   t^2+U^2}\right)-i \eta +\omega } 
\ee
\be
G_{aagg}=G_{aahh}\;\;\mbox{and}\;\;G_{bbgg}=G_{gghh}.
\ee
To construct $P_{[1/1]}[G_0]$ we can use \eqref{p11sigma} in the Dyson equation ${\bf G}={\bf G_0}+{\bf G_0}{\bf \Sigma} {\bf G}$,
where ${\bf G_0}={\bf G|_{U\to 0}}$. 
Provided with the exact functional dependence of the Green's function on the interaction parameter $U$
one can build ${\bf \Sigma_1}$ and ${\bf \Sigma_2}$ either via \eqref{Sigmafirstorder} and subsequent formulas or, equivalently, 
directly expanding in Taylor series the function ${\bf G}^{-1}-{\bf G_0}^{-1}$.
Upon appropriate rescaling of the infinitesimal parameter $\eta$, the two procedures lead to the same results:
\be
{\bf \Sigma}_1=
\begin{pmatrix}
 \frac{U}{2} & 0 & 0 & 0 \\
 0 & \frac{U}{2} & 0 & 0 \\
 0 & 0 & \frac{U}{2} & 0 \\
 0 & 0 & 0 & \frac{U}{2} \\
\end{pmatrix}
\;\;\;\;\;\;
{\bf \Sigma}_2=
\begin{pmatrix}
\frac{U^2}{4}\frac{1}{\omega-i \eta +3 t } & 0 & 0 & 0 \\
 0 & \frac{U^2}{4}\frac{1}{\omega+i \eta -3 t } & 0 & 0 \\
 0 & 0 & \frac{U^2}{4}\frac{1}{\omega-i \eta +3 t } & 0 \\
 0 & 0 & 0 & \frac{U^2}{4}\frac{1}{\omega+i \eta -3 t } \\
\end{pmatrix}
\ee
and
\begin{multline}
\left( {\bf \Sigma_1}^{-1}-{\bf \Sigma_1}^{-1}{\bf \Sigma_2 \Sigma_1}^{-1} \right)^{-1}=\\
=\begin{pmatrix}
\frac{U}{2}+\frac{U^2}{4 \left(\frac{1}{2} (6 t-U)-i \eta +\omega \right)} & 0 & 0 & 0
   \\
 0 & \frac{U}{2}+\frac{U^2}{4 \left(\frac{1}{2} (-6 t-U)+i \eta +\omega \right)} & 0 & 0
   \\
 0 & 0 & \frac{U}{2}+\frac{U^2}{4 \left(\frac{1}{2} (6 t-U)-i \eta +\omega \right)} & 0
   \\
 0 & 0 & 0 & \frac{U}{2}+\frac{U^2}{4 \left(\frac{1}{2} (-6 t-U)+i \eta +\omega \right)}
   \\\end{pmatrix}
\end{multline}
which turns out to be the exact self-energy.

\section{}\label{app:data}
We here report the values, rounded to the second digit,
of the average absolute deviation $\sigma$ plotted in fig.
\ref{conv_PN0},\ref{conv_P0N}, \ref{conv_PNN} and \ref{fig:P02vP11}.

\begin{table*}[ht]
    \includegraphics[scale=1]{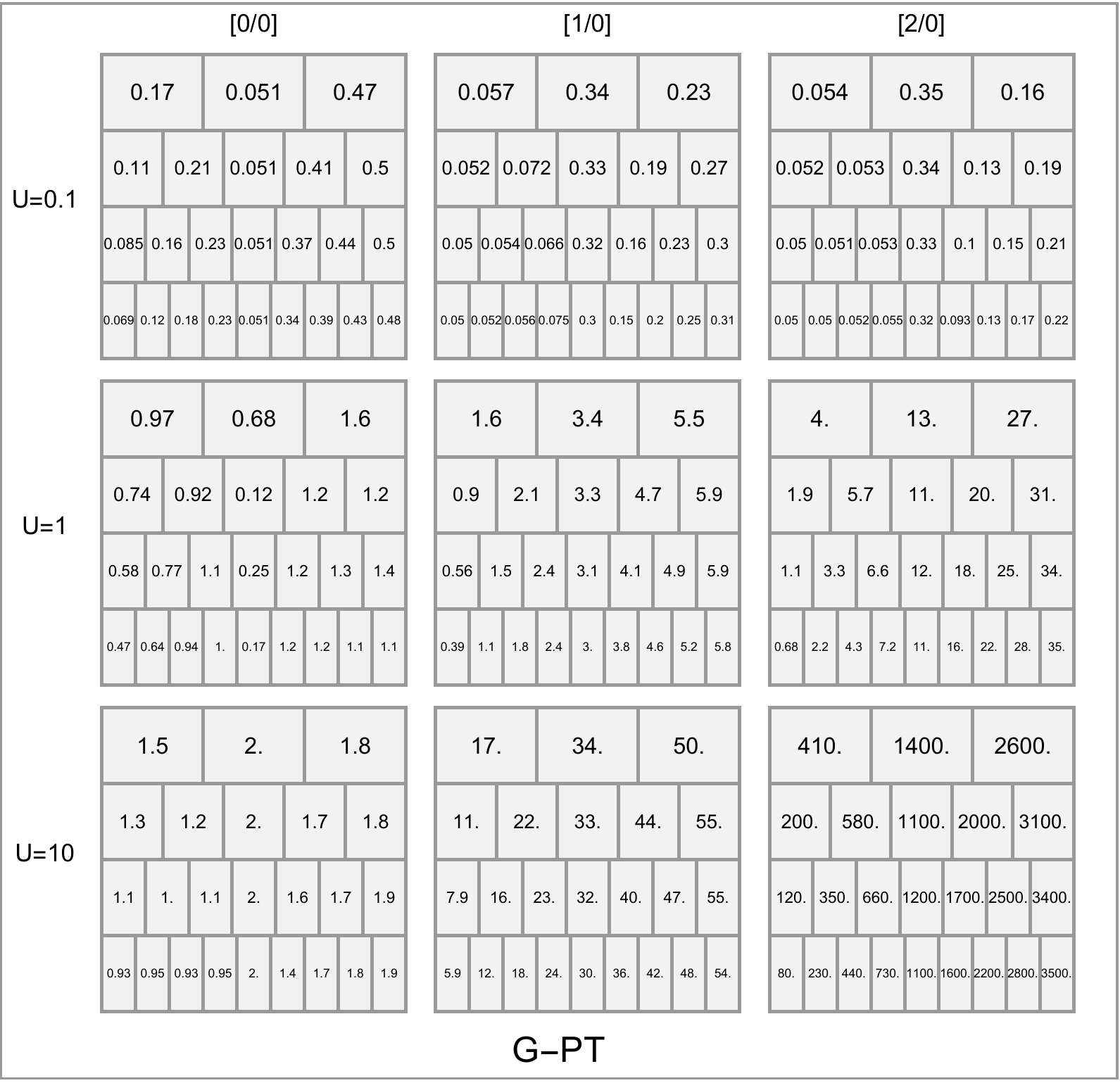}
    \caption{}
    %\label{conv_PN0}
\end{table*}
\begin{table*}[ht]
    \includegraphics[scale=1]{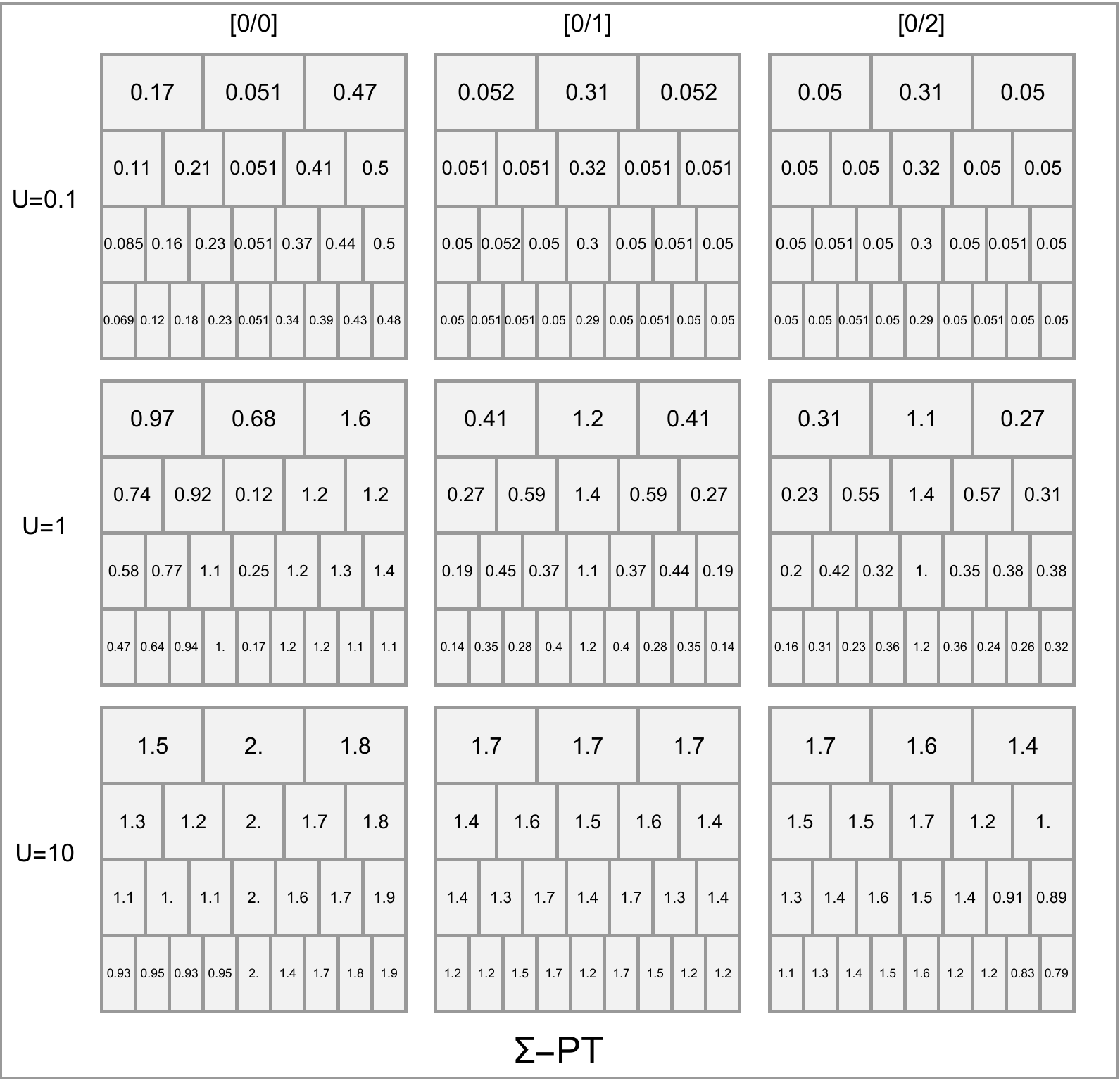}
    \caption{}
    %\label{conv_P0N}
\end{table*}
\begin{table*}[ht]
    \includegraphics[scale=1]{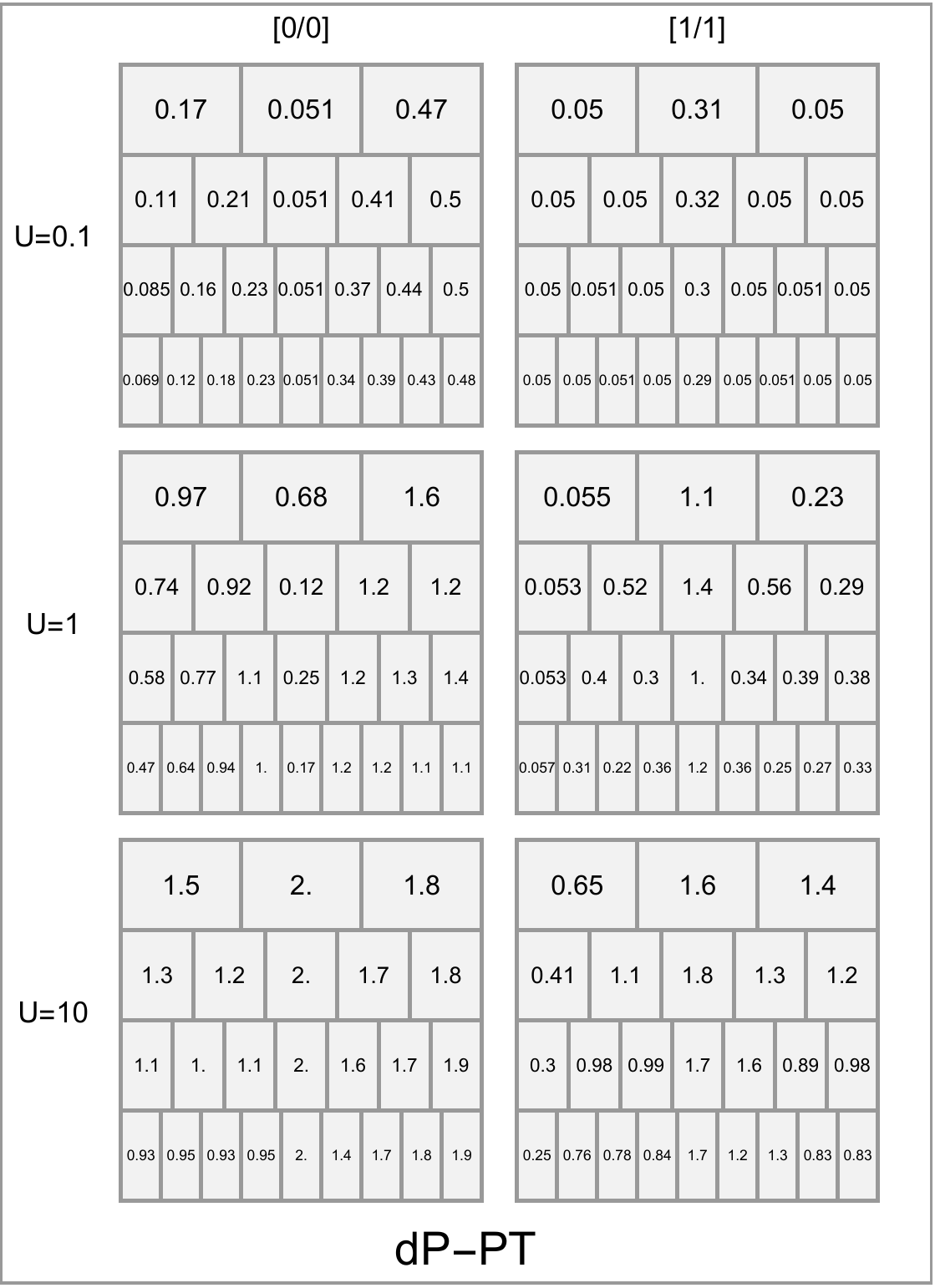}
    \caption{}
    %\label{conv_PNN}
\end{table*}
\begin{table*}[ht]
\includegraphics[scale=1]{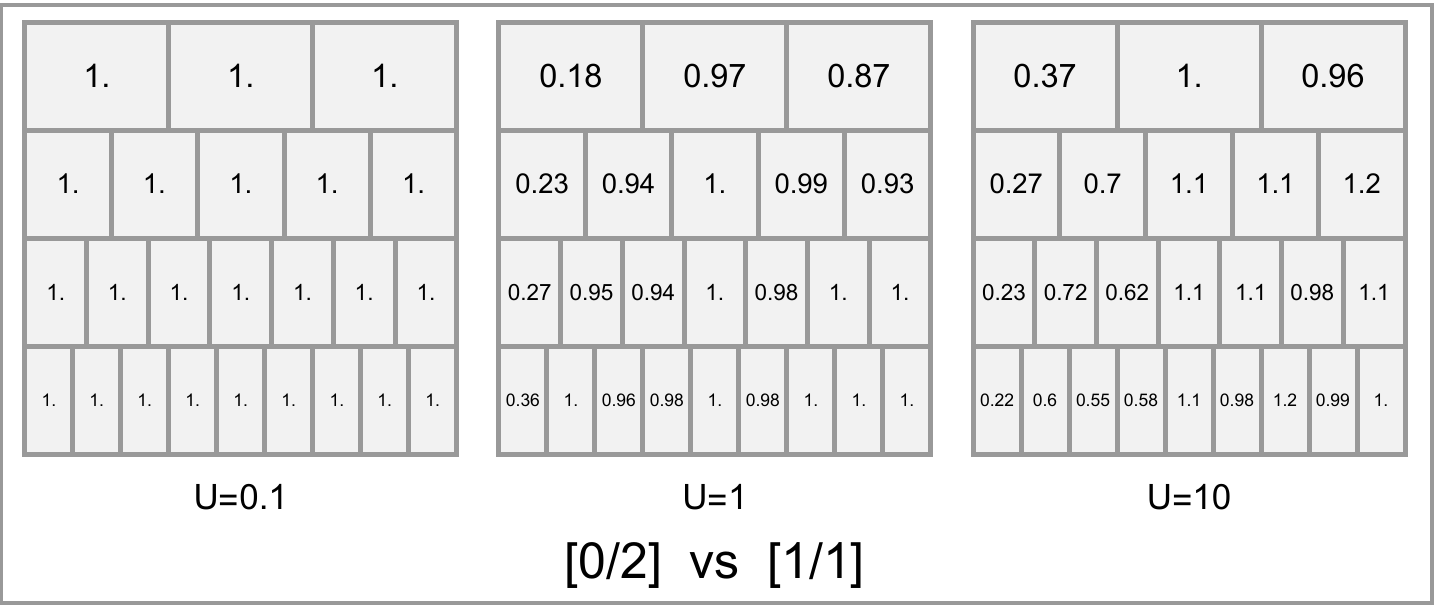}
\caption{}
%\label{fig:P02vP11}
\end{table*}

\bibliography{bibliography}
\end{document}